\documentclass[useAMS,usenatbib]{mn2e}

\usepackage{graphicx}
\usepackage{gensymb}
\usepackage{amsmath}
\usepackage{amssymb}
\usepackage{subfig}
\usepackage{rotating}
\usepackage{times}

\title[Misaligned streamers around a galactic centre BH]{Misaligned streamers around a galactic centre black hole
  from a single cloud's infall}
\author[W.~E.~Lucas et al.]{W.E.~Lucas$^1$\thanks{E-mail: 
  wel2@st-andrews.ac.uk}, I.A.~Bonnell$^1$, M.B.~Davies$^2$ \& W.K.M.~Rice$^3$ \\
  $^1$ SUPA, School of Physics \& Astronomy, University of
    St Andrews, North Haugh, St Andrews, Fife KY16 9SS, United Kingdom \\
  $^2$ Lund Observatory, Department of Astronomy and Theoretical Physics, 
    Box 43, SE-221 00 Lund, Sweden \\
  $^3$ SUPA, Institute for Astronomy, University of
    Edinburgh, Blackford Hill, Edinburgh EH9 3HJ, United Kingdom}
  
\begin{document}

\maketitle

\begin{abstract}  
  We follow the near radial infall of a prolate cloud onto a $4\times 10^6 M_\odot$ supermassive
  black hole in the Galactic Centre using smoothed particle hydrodynamics (SPH). We show that a 
  prolate cloud oriented perpendicular to its orbital plane naturally produces a spread in angular 
  momenta in the gas which can translate into misaligned discs as is seen in the young stars
  orbiting Sagittarius A*. A turbulent or otherwise highly structured cloud is necessary to avoid 
  cancelling too much angular momentum through shocks at closest approach. Our standard model of a
  $2\times 10^4 M_\odot$ gas cloud brought about the formation of a disc within $0.3\,\rmn{pc}$ from
  the black hole and a larger, misaligned streamer at $0.5\,\rmn{pc}$. A total of 
  $1.5\times 10^4 M_\odot$ of gas formed these structures. Our exploration of the simulation 
  parameter space showed that when star formation occurred, it resulted
  in top-heavy IMFs with stars on eccentric orbits with semi-major axes $0.02$ to 
  $0.3\,\rmn{pc}$ and inclinations following the gas discs and streamers. We suggest 
  that the single event of an infalling prolate cloud can explain the occurrence of multiple
  misaligned discs of young stars.
\end{abstract}

\section{Introduction} \label{s:intro}
  \citet{KR91} first discovered twelve definite massive O and Wolf--Rayet (W--R) stars orbiting our
  Galaxy's central black hole (BH) Sagittarius A* (Sgr A*). \citet{KR95} raised this to twenty-four,
  and more recently \citet{BA10} counted 177. These stars are intriguing for their unusual 
  situation. They appear to be coeval, having formed roughly $6\,\rmn{Myr}$ ago \citep{PA06}. All 
  orbit at distances between roughly $0.05$ and $0.5\,\rmn{pc}$ of Sgr A*. Further astrometric and
  spectroscopic measurements have confirmed that these stars form at least one eccentric 
  ($e = 0.36 \pm 0.06$) disc- or ring-like structure (\citealt{GE96}; \citealt{GE03}; 
  \citealt{PA06}; \citealt{LU09}; \citealt{BA09}; \citealt{BA10}) and fit a top-heavy IMF (eg. 
  $\rmn{d}N \propto m^{-0.45\pm0.3}\rmn{d}m$, \citealt{BA10}). Interestingly, a reasonable subset of
  the stars rotate oppositely in the plane of the sky. \citet{PA06} and \citet{BA09} concluded that
  these stars form a secondary stellar disc or streamer, although its existance is statistically 
  less certain \citep{LU09}.
  
  It seems likely that these stars were formed in a single event, though their unusual location and
  configuration complicate matters. \citet{GE01} suggested that a massive cluster similar to Arches
  or the Quintuplet could infall to the BH via dynamical friction with the background stars. The
  remnant cluster core would be deposited there. However, timescale arguments make this difficult 
  (see \citealt{HM03} and the discussions in \citealt{AL05} and \citealt*{GE10}). Another scenario 
  is that of the stars' formation \emph{in situ} within an accretion disc around the BH 
  \citep{LB03}. While `normal' star formation within a cloud so close to the BH would be strongly
  inhibited by tidal forces, a disc formed from such a cloud's tidal shearing around the BH may
  become self-gravitating to such a degree that fragmentation may occur (\citealt{KS80}; 
  \citealt{SB89}; \citealt{CZ99}; \citealt{GO03}; \citealt{LB03}; \citealt{NC05}; \citealt*{NCS07}). 
  
  In agreement, numerical simulations have shown that portions of an infalling cloud can be captured
  by a BH to form an eccentric disc that fragments to form a stellar disc population 
  (\citealt{BR08}; \citealt{HN09}; \citealt{AL11}; \citealt{MA12}). \citet{BR08} found that
  heating of the gas as it was compressed around the BH raised the Jeans mass, leading to the
  formation of a population of high mass stars in inner regions around the BH. Lower mass stars
  formed farther out in a separate population. \citet{HN09} found that a collision between two
  clouds close to Sgr A* could be responsible for the formation of multiple discs, though this would
  require two clouds to simultaneously enter the innermost Galactic Centre (GC). Two populations of 
  stars formed in the same manner as in \citet{BR08}.
  
  In this paper we use smoothed particle hydrodynamics to examine a model similar to that of 
  \citet{BR08}, requiring only one cloud on infall towards a BH -- a simpler and perhaps more
  frequent event.  
  The cloud's geometry is prolate to the orbital plane, generating a large spread in
  angular momentum when on a highly radial orbit as gas flows around the BH from different
  directions. This naturally leads to the formation of multiple misaligned gas structures orbiting 
  the central BH.

\section{Numerical method} \label{s:numerical}
  \subsection{Smoothed particle hydrodynamics} \label{ss:sph}
    Smoothed particle hydrodynamics (SPH) is a Lagrangian hydrodynamics formalism in which particles
    are used to represent portions of the total gas mass. A good description of the formalism is 
    provided by \citet{MO92}. We used the code of \citet*{BBP95}, itself derived from an earlier SPH
    code (\citealt{BE90}; \citealt{BEA90}). This code includes individual particle timesteps 
    (eg. \citealt{HK89}) and is parallelized using OpenMP. Sink particles, which are the SPH
    representations of (proto-)stars, were implemented following \citet{BBP95}. 
    
    Radiative transfer of energy was approximated with the hybrid method of \citet{FO09}. 
    Standard
    SPH shock heating and $p\,\rmn{d}V$ work were combined with the polytropic cooling technique of
    \citet{ST07} and the flux-limited diffusion method of \citet{MA07}. Polytropic cooling alone was
    used by \citet{BR08} and allowed energy transfer with an external radiation background. This
    took the form of cooling when the gas had a higher temperature than the background, and heating 
    when it was lower. The radiative diffusion used here further 
    improved the approximation by allowing energy to flow from hot to cold gas, the flux-limiter
    regulating the flow appropriately between the optically thick and thin limits. We used a
    background temperature of $T_\rmn{bg} = 100\,\rmn{K}$, close to the values observed in GC
    molecular and atomic gas (\citealt{JA93}; \citep{MPEA97}; \citealt{CH05}). Once a particle's
    time derivative of the internal energy had been calculated, it was semi-implicitly integrated
    as described by \citet{FO09}, following \citet{ST07}. In contrast to both this work and
    \citet{BR08}, the cooling of \citet{HN09} was determined via a parametric relation with the
    dynamical timescale.

    Particles were integrated in thirty timestep bins, each with a power of two division of the
    maximum timestep $\Delta t_\rmn{max} = 236\,\rmn{yrs}$, or $5 \times 10^{-4}$ in code units (one
    code unit was equivalent to $1.4874 \times 10^{13}\,\rmn{s}$). The smallest timestep a particle
    could be integrated on was $\Delta t_\rmn{max}/2^{29} = 14\,\rmn{s}$, though typically few ever
    required a timestep shorter than $\Delta t_\rmn{max}/2^{20} = 2\,\rmn{hours}$. While a large BH
    accretion radius (see 
    Section~\ref{ss:sphsinks}) delayed it, the accretion to the disc of cloud material over time
    eventually caused particle timesteps to fall below the minimum, ending the simulation. To keep
    the simulation running would have required reducing the maximum timestep or increasing the
    number of bins: either would have unreasonably increased the integration time.
    
    Gas in the simulation experienced the potential from the nuclear stellar cluster in addition
    to the BH and its own self-gravity. To obtain the enclosed mass as a function of Galactocentric 
    radius $R$, we integrated over volume the stellar number density of \citet{ME10},
    who constructed a model using observations by \citet{SEA07} and \citet*{BSE09}. This number
    density is given as
    \begin{equation}\label{eq:densitylaw}
      n(R) = n_0 \left( \frac{R}{R_0}\right)^{-\gamma_i} \left [ 1+\left( 
        \frac{R}{R_0}\right)^{\alpha} \right]^{(\gamma_i - \gamma)/\alpha}.
    \end{equation}
    To tabulate the masses we used their best fit parameters of $\gamma_i = -1.0$, $\gamma = 1.8$ 
    and $\alpha = 4.0$, $R_0 = 0.21\,\rmn{pc}$, and normalized with the observed enclosed mass at 
    $R=1\,\rmn{pc}$ of $\approx 1 \times 10^6 M_\odot$ \citep*{SME09}.
    
  \subsection{Sinks in SPH} \label{ss:sphsinks}
    Sinks are N-body particles that only experience gravitational forces, but can accrete
    nearby gas particles. Dynamic sink creation was used to model star formation. Following 
    \citet{BBP95}, sinks were created to replace a bound, collapsing region of gas. 
    Such a region 
    was required to be smaller than the accretion radius of the sink which would replace it (see 
    next paragraph), $10^{-3}\,\rmn{pc}$. It was also required to exceed a critical density. In 
    those runs which used clouds with masses $10^4 M_\odot$ and $2 \times 10^4 M_\odot$, it was 
    $10^{-11}\,\rmn{g}\,\rmn{cm}^{-3}$, while for the clouds of $10^5 M_\odot$, which were less 
    dense initially, it was $6.4\times 10^{-12}\,\rmn{g}\,\rmn{cm}^{-3}$. These are of order ten
    million times the initial cloud densities. They also approximate the tidal density at the BH's
    outer accretion radius, helping prevent sink formation where tidal disruption is likely.
    Once a region fulfilling the above requirements was identified, further checks were made to 
    ensure that the region was both bound and collapsing, as described by \citet{BBP95}. The initial
    mass for sink particles was 
    approximately fifty times the gas particle mass. For clouds with $2 \times 10^4 M_\odot$, this 
    was $0.3 M_\odot$; for $10^5 M_\odot$, it was $1.5 M_\odot$. Taken together, these tests ensured
    that a real and resolved star formation event was taking place.
    
    Accretion to a sink from within an outer accretion radius $r_\rmn{acc,out}$ was only allowed if
    the gas was bound to the sink, and more tightly than to any other sink, and if its angular 
    momentum around it was small enough to allow it to enter a circular orbit at $r_\rmn{acc,out}$.
    Conversely, all gas was accreted once it fell within an inner accretion radius $r_\rmn{acc,in}$.
    We used $2.5\times 10^{-4}\,\rmn{pc}\approx 50\,\rmn{AU}$ for both the stars and the BH.
    $r_\rmn{acc,out} = 10^{-3}\,\rmn{pc} \approx 200\,\rmn{AU}$ was used for sink particles
    representing stars, the same as the critical sink creation radius. The outer accretion radius 
    was chosen to approximate the size of a protostellar disc; therefore a sink particle may be
    taken to represent an unresolved star-disc system. This radius was chosen as an acceptable 
    tradeoff between simulation runtime, decreased with increasing $r_\rmn{acc}$, and sink 
    resolution, improved with decreased $r_\rmn{acc}$.
    
    We represented the black hole with a sink particle of mass $4\times 10^6 M_{\sun}$ 
    \citep{GH08}.
    In order to avoid relativistic effects and huge forces on gas particles, we increased
    its outer accretion radius to $0.02\,\rmn{pc}$. At this distance from the BH, the Keplerian
    circular speed was $0.3$ per cent of the speed of light. The inner accretion radius was the same
    as that used for dynamically created sinks.
    
    It is important to note that dynamically-created sink particles do not represent true stars. 
    Rather, they consist of an unresolved accretion disc and (proto)star that exist within the 
    accretion radius. Likewise, the large accretion disc around the BH would in reality extend
    inside the outer accretion radius.
    
  \subsection{Physical setup}
  
    \begin{figure} \centering
    \includegraphics[scale=0.13]{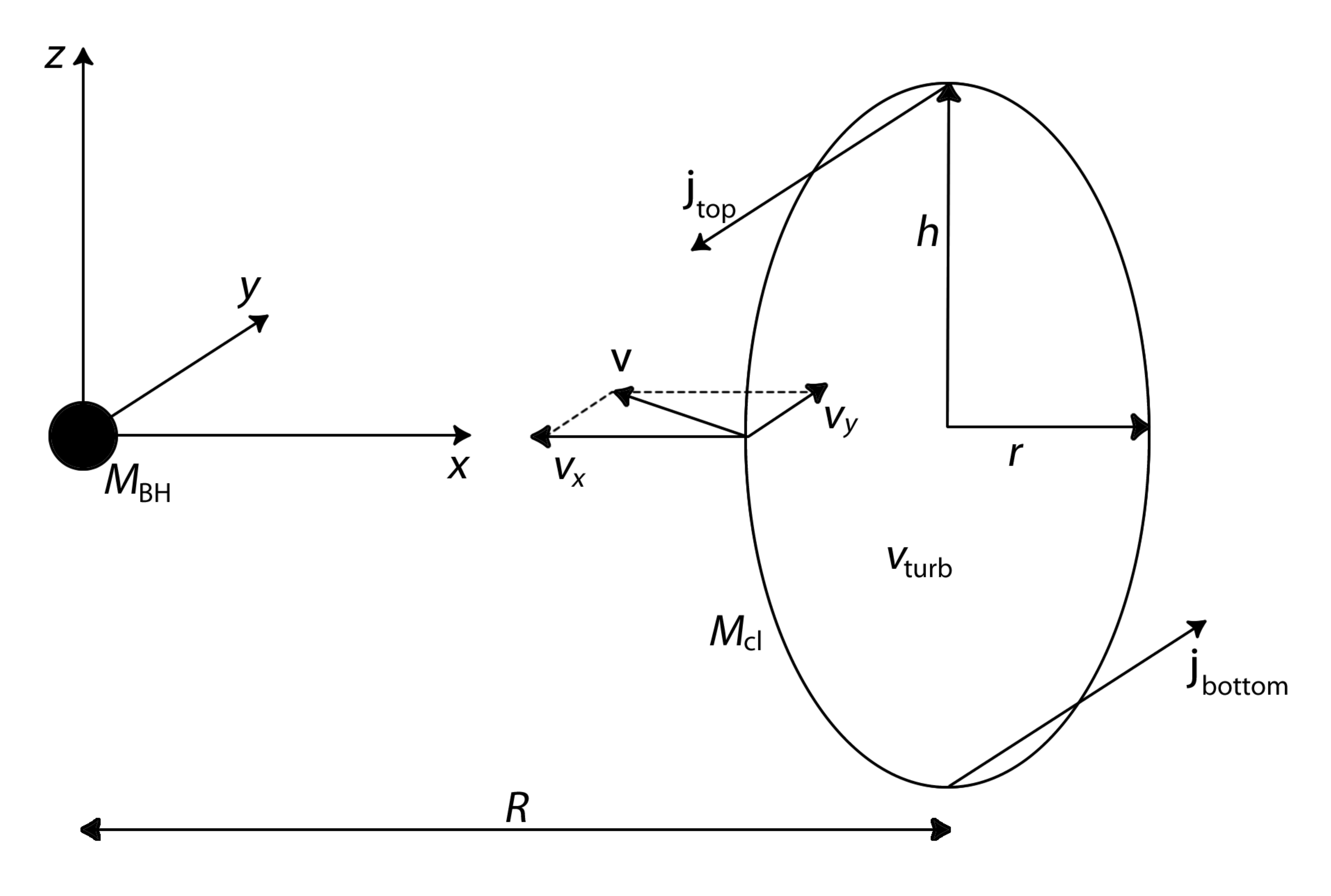}
    \caption{Here we show the initial simulation geometry, comprising two objects:
      a gas cloud of mass $M_\rmn{cl}$ and a black hole (BH), represented by a sink particle of
      mass $M_\rmn{BH}$. The cloud was positioned such that its centre lay on the $x$-axis at a 
      distance $R$ from the BH which lay at the origin. It had a semi-minor axis of length $r$ and
      semi-major axis of length $h$, which was aligned with the $z$-axis. The cloud was given
      an initial velocity $\mathbf{v}$ which lay in the $xy$-plane, giving it an infall velocity
      $v_x$ and a tangential velocity $v_y$. This ensures that specific angular momentum 
      $\mathbf{j}$ is spread over a large angle between particles at the top $\mathbf{j}_\rmn{top}$
      and bottom $\mathbf{j}_\rmn{bottom}$ of the cloud.}
    \label{infall_geom}
    \end{figure}
    
    \begin{table*}
    \caption{Run names and their initial conditions. $M_\rmn{cl}$ was the cloud's mass, $r$ was
      the length of its semi-minor axis and $h$ its semi-major axis. The cloud was given an initial
      velocity $\mathbf{v}_\rmn{cl}$; integrating the system by replacing the cloud with a point
      mass gave the minimum distance from the BH $R_\rmn{min}$ during the pass. Other changes 
      are noted. $t_\rmn{end}$ is the final simulation time. $\Delta M_\rmn{BH}$ is the mass
      accreted to the BH during the simulation. Finally we have the number of sinks formed by the
      run's end $N_\rmn{sink}$ and the total mass in sinks $M_\rmn{sink,tot}$ (excluding the BH).
      Note these values are not directly comparable as the simulations finished at different
      times.}
    \label{table:models}
    \begin{tabular}{@{}lcccccccccc}
      \hline
      Run name & $M_\rmn{cl}/M_\odot$ & $r (\rmn{pc})$ & $h (\rmn{pc})$ & $\mathbf{v}_\rmn{cl} (\rmn{km}\,\rmn{s}^{-1})$
       & $R_\rmn{min} (\rmn{pc})$ & Notes & $t_\rmn{end} (yrs)$ & $\Delta M_\rmn{BH} (M_\odot)$ & $N_\rmn{sink}$ & $M_\rmn{sink,tot} (M_\odot)$ \\
      \hline
      %standard_menc
      A & $2\times 10^4$ & $0.4$ & $1.0$
        & $(-41.5,10.4,0)$ & $0.022$ & - & $28300$ & $1280$ & $14$ & $21.9$ \\
      %standard_menc_new
      A re-run & $2\times 10^4$ & $0.4$ & $1.0$
        & $(-41.5,10.4,0)$ & $0.022$ & - & $30890$ & $4290$ & $10$ & $9.00$ \\
      %nocluster
      B & $2\times 10^4$ & $0.4$ & $1.0$
        & $(-41.5,10.4,0)$ & $0.028$ & No cluster & $32540$ & $553$ & $19$ & $32.0$ \\
      %hires
      C & $2\times 10^4$ & $0.4$ & $1.0$
        & $(-41.5,10.4,0)$ & $0.022$ & $10^7$ particles & $26650$ & $3870$ & $5$ & $3.93$ \\
      %angmom2_menc
      D & $1\times 10^4$ & $0.4$ & $1.0$
        & $(-41.5,20.7,0)$ & $0.064$ & - & $37020$ & $45.8$ & $25$ & $59.2$ \\
      %angmom_menc
      E & $1\times 10^4$ & $0.4$ & $1.0$
        & $(-41.5,41.5,0)$ & $0.163$ & - & $36650$ & $0.00$ & $109$ & $375$ \\   
      %density_menc
      F & $1\times 10^5$ & $1.0$ & $2.5$
        & $(-41.5,10.4,0)$ & $0.022$ & - & $25000$ & $8940$ & $64$ & $741$ \\
      %turb_menc
      G & $2\times 10^4 $ & $0.4$ & $1.0$
        & $(-41.5,10.4,0)$ & $0.022$ & Alt. turbulence & $23110$ & $726$ & $17$ & $62.6$ \\
      %infall150_tang5
      H5 & $1\times 10^4 $ & $0.4$ & $1.0$
        & $(-150,5,0)$ & $0.006$ & - & $15800$ & $5230$ & $1$ & $1.95$ \\
      %infall150_tang10
      H10 & $1\times 10^4 $ & $0.4$ & $1.0$
        & $(-150,10,0)$ & $0.021$ & - & $15560$ & $2830$ & $0$ & $0.00$ \\
      %infall150_tang20
      H20 & $1\times 10^4 $ & $0.4$ & $1.0$
        & $(-150,20,0)$ & $0.060$ & -& $16740$ & $348$ & $0$ & $0.00$ \\
      %infall150_tang40
      H40 & $1\times 10^4 $ & $0.4$ & $1.0$
        & $(-150,40,0)$ & $0.153$ & - & $96220$ & $5.59$ & $0$ & $0.00$ \\
      %infall150_tang5_density
      I5 & $1\times 10^5$ & $1.0$ & $2.5$
        & $(-150,5,0)$ & $0.006$ & - & $25470$ & $38200$ & $1$ & $1.59$ \\
      %infall150_tang10_density
      I10 & $1\times 10^5$ & $1.0$ & $2.5$
        & $(-150,10,0)$ & $0.020$ & - & $27360$ & $37100$ & $0$ & $0.00$ \\
      %infall150_tang20_density
      I20 & $1\times 10^5$ & $1.0$ & $2.5$
        & $(-150,20,0)$ & $0.060$ & - & $18870$ & $20000$ & $68$ & $1190$ \\
      %infall150_tang40_density
      I40 & $1\times 10^5$ & $1.0$ & $2.5$
        & $(-150,40,0)$ & $0.152$ & - & $20750$ & $2410$ & $10$ & $453$ \\
      \hline
    \end{tabular}
  \end{table*} 
    
    The simulations followed the infall of a single prolate ellipsoidal cloud. The parameters for
    each are listed in Table~\ref{table:models}. The cloud had had semi-minor 
    axis $r$ and semi-major axis $h$, and was positioned such that its centre lay at a distance 
    $R = 3\,\rmn{pc}$ from the BH which was placed at the origin. Its major axis was parallel to the
    $z$-axis, while it was given an initial velocity $\mathbf{v}$ in the $xy$-plane. The combination
    of the cloud's shape and its highly radial orbit bestowed a large spread in angular momentum.
    The simplicity of this setup, depicted in Figure~\ref{infall_geom}, is its principle 
    attractive feature.
    
    We also applied to the cloud an initial turbulent velocity field which provided support
    against its self-gravity and also generated structure -- this proved to be crucial in
    the retention of the gas's angular momentum about the BH, and will be discussed later in
    Section~\ref{ss:turb}. The method used was that described by \citet{DNP95} and \citet{DBC05}. 
    The velocity field was drawn from a power spectrum
    \begin{equation}
      P(k) \equiv \langle |v_k|^2  \rangle \propto k^{-n}
    \end{equation}
    with $n=3.5$, by randomly sampling a vector potential, making the field initially 
    divergenceless. Gas particles
    velocities were interpolated from the output grid, and then scaled in all our simulations to 
    give a ratio of turbulent kinetic to gravitational energies $|E_\rmn{kin}/E_\rmn{grav}| =  1.5$
    
    The initial conditions of our simulations are shown in Table~\ref{table:models}. The simulation
    we call Run~A is used to demonstrate the model. It used
    $M_\rmn{cl} = 2\times 10^4 M_\odot$, $r = 0.4\,\rmn{pc}$, $h = 1.0\,\rmn{pc}$ and
    $\mathbf{v}_\rmn{cl} = (-41.5,10.4,0)\,\rmn{km}\,\rmn{s}^{-1}$. We used $3141792$ particles to
    represent the gas.
    
    We motivate our infall velocities from observations \citep{TEA11} which have found little gas 
    with line-of-sight speeds beyond 
    $75\,\rmn{km}\,\rmn{s}^{-1}$ within $\sim 35\,\rmn{pc}$ projected distance from the GC. Thus we
    initially used a low infall velocity of $v_x = -41.5\,\rmn{km}\,\rmn{s}^{-1}$. Calculations of
    infall from rest between $5$ and $20\,\rmn{pc}$ indicated that by $R = 3\,\rmn{pc}$, a test
    particle would be moving between $100$ and $200\,\rmn{km}\,\rmn{s}^{-1}$ due to the BH and 
    cluster. As a result, we performed several more simulations using 
    $v_x=-150\,\rmn{km}\,\rmn{s}^{-1}$.
    
    Here we briefly discuss Run~B and C as they were performed as tests of the setup.
    Run~B did not include the stellar cluster. As expected, the missing extended
    potential caused the gas particles to follow trajectories more similar to closed Keplerian 
    orbits than those in Run~A. Otherwise they were very similar, demonstrating the dominance of
    the BH potential in the central parsec.
    
    Run~C was a resolution test, performed using $10$ million particles. In order to ensure 
    it ran 
    long enough over a timescale similar to the other simulations, we relaxed the BH accretion 
    rules, allowing it to accrete all gas within $0.02\,\rmn{pc}$ without test. Although this only
    affected the innermost regions of the simulation, of secondary interest to the large scale
    gas dynamics, we re-ran Run~A with the same accretion rules to allow a direct comparison.
    Run~C's
    improved mass resolution of $1.91\times 10^{-3} M_\odot$ caused sinks to form at masses roughly
    three times smaller than in the other simulations. Despite this, by the time of comparison
    five sinks had formed and accreted to masses of $\approx 1 M_\odot$ in both simulations.
    The gas distribution was also highly comparable, with the only change being slightly higher
    densities close to the BH accretion radius. Consequently, the BH had accreted slightly more in
    Run~C at $3870 M_\odot$ compared to $3550 M_\odot$ in the re-run of A. As such we accept the 
    set of lower resolution simulations as being reasonably well resolved. We have included 
    the re-run of A in Table~\ref{table:models}, but do not discuss it further.

\section{Gas dynamics} \label{s:gasdynam}
  
  \subsection{Formation of a misaligned streamer} \label{ss:infallproc}
    Run~A provides a good demonstration of the sequence of events during infall and the formation of
    a misaligned structure. The cloud began its infall with uniform density, but the 
    turbulence quickly generated structure and removed the symmetry between top and
    bottom. As the cloud drew closer, the tidal forces from the black
    hole began to pull the cloud apart. The tangential velocity given to the
    cloud was enough that its centre of mass passed to the side of the BH, giving enough angular
    momentum to prevent the majority of the cloud from being accreted.
    
    \begin{figure*} \centering
      \includegraphics[scale=2.2]{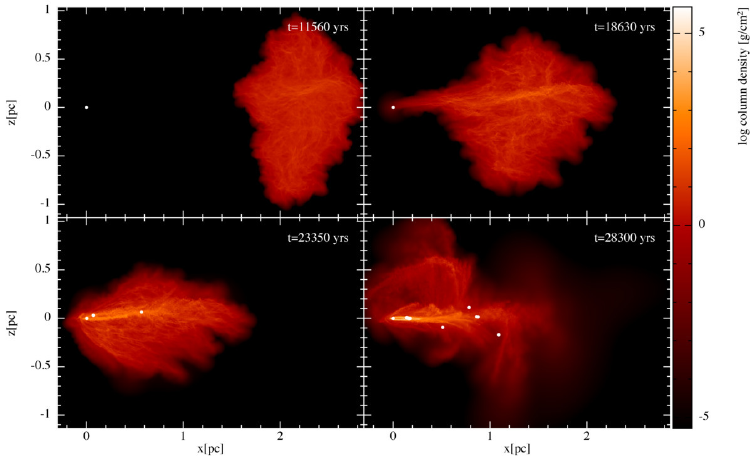}
      \caption{Column density in $xz$ of Run~A's cloud at four times. The turbulence in the cloud 
        generated enough structure that by the time it reached the black hole the flow around it
        was highly asymmetric, and had actually formed several sink particles. These though were on
        almost radial orbits and had semi-major axes of between $1$ and $3\,\rmn{pc}$. The BH and
        dynamically created sink particles are shown as white dots. As a sink, the BH was allowed to
        move, and so was
        accelerated towards the cloud. However, its huge mass meant its motion is not visible on the
        scales shown here. The misaligned streamer can be seen to sweep up from below around the BH.
        It was however never dense enough to form sinks.}
      \label{ColDens_standard_menc_GC2D050_GC2D080_GC2D100_GC2D121_zx}
    \end{figure*}
    
    During its pass of the BH, the cloud was tidally sheared. some of its mass being captured and
    forming a disc on the scale of $\sim 0.1\,\rmn{pc}$. Compression of the gas as it flowed around
    the BH heated it to a maximum of $6000\,\rmn{K}$, before it expanded and was allowed to cool 
    again. At closest approach, gas passed over and under the BH, which, combined with the 
    tangential motions, produced streamers of gas with very different angular momenta.
    Gas which had accumulated in a central overdensity formed a small 
    dense disc due to its comparitively low angular momentum. Another overdensity which had formed 
    in the bottom (negative $z$ region) of the cloud swept up to form a large streamer at 
    $60 \degree$ to the dense inner disc. By the end of the simulation, at 
    $t=28300\,\rmn{yrs}$, $15500M_\odot$ of gas was bound within a distance of $1\,\rmn{pc}$ from 
    the BH, which had itself accreted $1280M_\odot$. Star formation in the cloud and disc produced 
    14 sinks (see Section~\ref{s:sf}).The process of infall can be seen in 
    Figure~\ref{ColDens_standard_menc_GC2D050_GC2D080_GC2D100_GC2D121_zx}.
    
\begin{figure} \centering
      \includegraphics[scale=1.2]{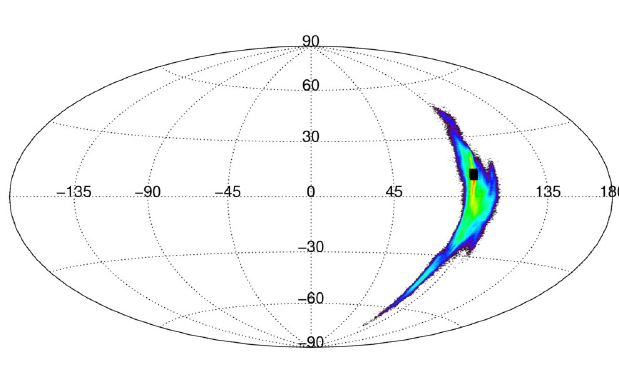} \\
      \includegraphics[scale=1.2]{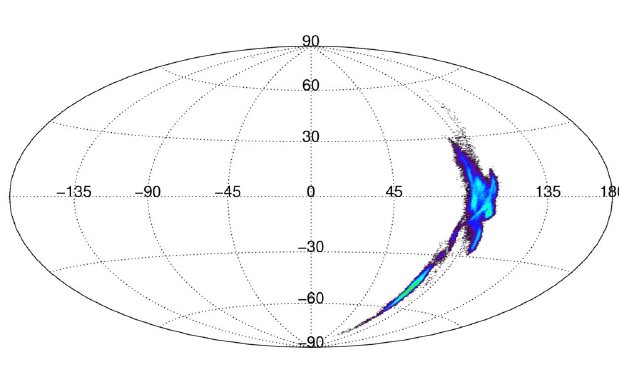}
      \caption{Angular momentum orientations on the sky for the gas particles in Run~A at the run's
        end at $t=28300\,\rmn{yrs}$ in a Hammer projection. The first and second plot 
        respectively show gas with semi-major axis less than and greater than $0.7\,\rmn{pc}$. The 
        vectors' components were calculated in ($\theta$, $\phi$) polar and azimuthal
        angles. These have been rotated such that the direction given by $(0,90)\degree$ points in
        positive $z$ (see Figure~\ref{infall_geom}) and corresponds to anticlockwise rotation in the
        $xy$-plane when looking down the $z$-axis. The direction $(0,0)\degree$ points along the 
        $x$-axis, and $(-90,0)\degree$ along $y$. Particles were then placed in equal-area bins. The log of the number of 
        particles in each bin has been plotted proportional to its shade, with both plots using the
        same colour scale. The large overall spread in angular momentum is apparent in both plots.
        The disc and streamer are seen as peaks in both plots located at $(10,90)\degree$ 
        and $(-50,75)\degree$, separated by $60\degree$.
        In the first plot the highest densities are located at the disc orientation. The black 
        squares, representing sink particles, align very well with one another and the disc. For 
        semi-major axes greater than $0.7\,\rmn{pc}$ the disc and streamer are much more distinct, 
        and the highest densities occur in the region of the streamer orientation.}
      \label{standard_menc_GC2D121_angmom}
    \end{figure}

    Our decision to use a cloud prolate to the orbital plane was responsible for the
    creation of the misaligned gas streamer. Visualising the spread in angular momentum direction
    on the sky reveals the presence of the disc and streamer -- 
    Figure~\ref{standard_menc_GC2D121_angmom}, 
    shows this in two plots, for gas with semi-major axis less and greater than 
    $0.7\,\rmn{pc}$. There are two peaks in the density of particles' angular momentum orientations
    on the sky representing both the central disc and the streamer. The density is noticeably 
    highest in the region inhabited by the disc for that gas on small orbits. In the second plot
    for gas on larger orbits, higher densities occur for the streamer. The overall distribution is
    visibly asymmetric about the $\theta = 0\degree$ line -- this asymmetry was present in the 
    initial distribution. As discussed in Section~\ref{ss:turb}, a more symmetric distribution
    results in the destruction of the misaligned streamers.
    
\begin{figure} \centering
      \includegraphics[scale=1.3]{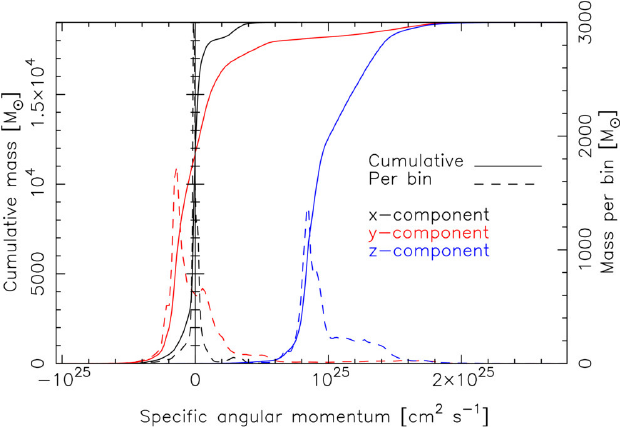}
      \caption{At $t=28300\,\rmn{yrs}$ in Run~A, the specific angular momentum for each gas 
        particle around the BH was calculated. The particles were binned by each component of the
        angular momentum; here we show the cumulative mass across these bins, as well as the
        individual values. The black line, showing the $x$-component $l_x$, is comparitively 
        symmetric about $0$ as $x$ was the infall axis. On the other hand, no gas has negative 
        $l_z$. This is due to the initial tangential speed $v_y$
        given to the cloud, which gave rotation around the $z$-axis and thus positive angular
        momentum. The most interesting is the $y$-component. Most mass has negative values, with
        the peak in bin mass also seen at negative angular momentum. 
        This is the disc, which formed
        from an overdensity slightly above the $xy$-plane and so possessed rotation around the
        $y$-axis.
        Gas misaligned with the disc, including the streamer, forms that part of the distribution
        extending to large positive $l_y$. It contains $\sim 2\times 10^3 M_\odot$.}
      \label{standard_menc_GC2D121_cumulative_mass_angmom}
    \end{figure}

    The asymmetry can also be seen in Figure~\ref{standard_menc_GC2D121_cumulative_mass_angmom}
    where gas at the end of Run~A has been binned by each component of its specific angular
    momentum. While the mass distribution of the $x$-component $l_x$ is symmetric, and that of $l_z$
    is offset due to the cloud's non-zero initial $v_y$, the $l_y$ distribution peaks at small 
    negative values. This is 
    the material originally within the top half of the cloud, much of which goes on to form the
    disc. The distribution then increases slowly to the $l_y$, forming a long but shallow tail of
    $\sim 2000 M_\odot$. The gas in the streamer is found in this tail.
    
\begin{figure} \centering
      \includegraphics[scale=1.3]{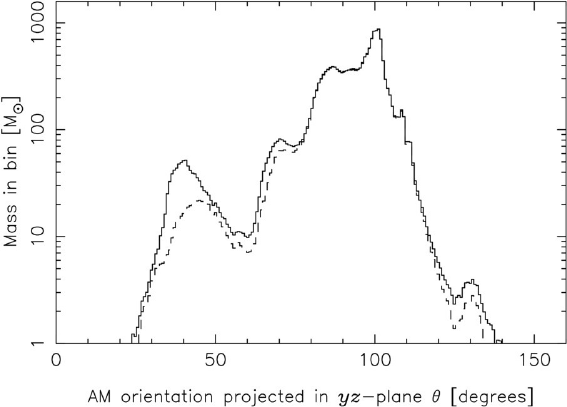}
      \caption{Here we show the gas in Run~A at $t=28300\,\rmn{yrs}$, binned by the angle between
        the angular momentum vector projected into the $yz$-plane and the $y$-axis, given by
        $\theta = \rmn{tan}(l_z/l_y)$. The solid line represents all gas in the simulation, while
        the dashed line shows only that which was bound to within one parsec; we have also given
        the bin values in logspace to better show the gas orbiting in planes misaligned with the
        disc. The disc makes up the large peak centred at $100\degree$. At $40\degree$ a smaller
        peak can be seen, making up the streamer. In contrast with the disc, half the gas in the
        streamer is bound to within one parsec. If we define the streamer as containing all the
        gas with $\theta \le 60\degree$, then its total mass is $1070 M_\odot$. The bound mass is
        $560 M_\odot$. Of the remaining $17650 M_\odot$, $17050 M_\odot$ is bound.
      }
      \label{standard_menc_GC2D121_angmom_angle_bins_log}
    \end{figure}

    The streamer mass may be further constrained by binning the gas by the angle between the 
    $y$-axis and its angular momentum vector projected into the $yz$-plane,
    $\theta = \rmn{tan}(l_z/l_y)$. This reveals a twin peaked distribution, as shown in 
    Figure~\ref{standard_menc_GC2D121_angmom_angle_bins_log}. The smaller peak, 
    representing the misaligned gas, was located $60\degree$ away from the larger peak, which is the
    disc, and contained $1070 M_\odot$. Only $560 M_\odot$ of this orbited within $1\,\rmn{pc}$ of
    the BH, the remainder on orbits with semi-major axes of up to $10\,\rmn{pc}$.
    
  \subsection{Importance of asymmetries} \label{ss:turb}    
    It is important for the infalling cloud to be structured if misaligned streamers are to form.
    If this is the case, the gas streams passing on either side of the BH do not shock completely
    and cancel their respective angular momenta. The turbulence used in Run~A 
    (see Figure~\ref{ColDens_standard_menc_GC2D050_GC2D080_GC2D100_GC2D121_zx}) created
    sufficient structure to accomplish this.
    
\begin{figure*} \centering
      \includegraphics[scale=2.2]{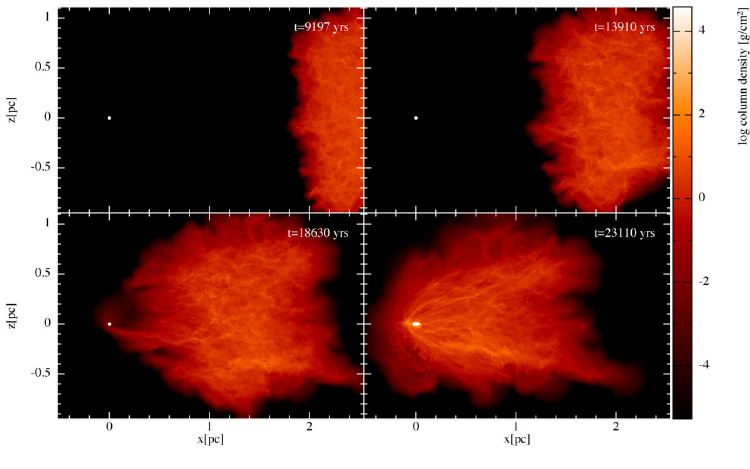}
      \caption{Column densities in the $xz$-plane throughout Run~G. The initial state was exactly
        the same as that in Run~A save that the turbulent velocity field was produced using
        different seed integers. This gave rise to a cloud lacking large overdense regions, and
        which was comparatively symmetric above and below the $xy$-plane. As can be seen, this led
        to the flows from these regions roughly mirroring one another and shocking on the other side
        of the BH. No misaligned structures formed. The cancellation of angular momentum via
        shocking also reduced the disc's size and eccentricity.}
      \label{ColDens_turb_menc_GC2D040_GC2D060_GC2D080_GC2E099_xz}
    \end{figure*}

    To explore the importance of a particular turbulent velocity field, we performed Run~G which 
    can be seen in Figure~\ref{ColDens_turb_menc_GC2D040_GC2D060_GC2D080_GC2E099_xz}. The 
    alternative realisation failed to produce significant asymmetries. Thus, 
    self-shocking of the flows occurred on reaching the BH. A disc formed from the central regions, 
    but the mirrored flows from the top and bottom met and shocked with one another. The disc
    itself was very small, with a radius $\approx 0.05\,\rmn{pc}$, and circular, again as a
    result of angular momentum cancellation.
    
\begin{figure} \centering
      \includegraphics[scale=1.2]{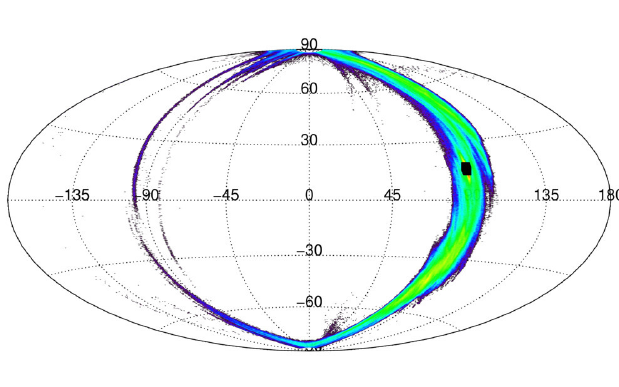}
      \caption{Density of angular momentum orientation in the sky for Run~G at $23100\,\rmn{yrs}$,
        constructed in the same way as in Figure~\ref{standard_menc_GC2D121_angmom},
        though here we include all the gas.
        The disc is present as the densest region, aligned with the black squares representing the
        direction of the sinks' angular momentum. In comparison to 
        Figure~\ref{standard_menc_GC2D121_angmom}, this distribution for gas is roughly symmetric 
        above and
        below the $\theta = 0\degree$ line, representing a large spread in the angular momentum of
        the infalling gas. The gas lying above the equator has total mass $9540 M_\odot$, while the
        counterrotating gas in the bottom hemisphere has a very similar $9610 M_\odot$. At the same 
        time, Run~A had values of $10420 M_\odot$ and $9215 M_\odot$, a difference of more than a 
        thousand solar masses. With the initial turbulent velocity field failing to generate
        significant structure, the misaligned gas flows met and shocked on the simulation 
        $xy$-plane, preventing the formation of a streamer.}
      \label{turb_menc_GC2E099_angmom}
    \end{figure}

    A comparison between the total angular momentum at two times, including that of gas accreted to
    the BH, is telling. Between the start and the end of Run~G, $20$ per cent of the total
    angular momentum magnitude
    was cancelled out. Over the same period, $12$ per cent of the angular momentum in Run~A was 
    cancelled out. A re-run of G using relaxed BH accretion rules progressed slightly further to 
    $24050\,\rmn{yrs}$ and experienced a further 
    $7$ per cent decrease. The plot of angular momentum distribution on the sky 
    (Figure~\ref{turb_menc_GC2E099_angmom}) shows symmetry above and below the equator which 
    indicates roughly equal gas flows orbiting in opposite directions around the BH.
    Calculations show that there is a $70 M_\odot$ difference in the mass of gas flowing
    in each direction in Run~G, while in Run~A at the same time the difference was $1205 M_\odot$.
    
    It is clear that for multiple structures to form, the various regions of the cloud which are to
    orbit the BH in such different planes must be able to retain their angular momentum, and it
    is self-shocking of the gas that acts to oppose to this. However, our use of turbulence was
    mainly to drive structure formation. Real molecular clouds are observed to be clumpy which would
    likely provide the asymmetry we require here.
    
  \subsection{Varying the cloud orbit} \label{ss:orbit}
    In other runs we varied the cloud's initial velocity $\mathbf{v}$ in order to examine the
    effect this had on the formation of the disc and any streamers. Run~A used $v_x = 
    -41.5\,\rmn{km}\,\rmn{s}^{-1}$ and a tangential speed of $v_y = 10.4\,\rmn{km}\,
    \rmn{s}^{-1}$. As such the cloud marginally engulfed the BH, though the majority of gas passed
    to one side. 
    
    Runs~D and E used higher tangential speeds of $v_y = 20.7\,\rmn{km}\,\rmn{s}^{-1}$ and
    $41.5\,\rmn{km}\,\rmn{s}^{-1}$. The large $v_y$
    meant that the cloud's trajectory in Run~D missed the BH; very little gas was captured to form
    a disc. With the high $v_y$, the large spread in angular momentum bestowed by the cloud's shape
    was compressed and the tidal arc formed during the pass around the BH was close to planar. The
    high angular momentum also meant that only small amounts of gas had been able to accrete onto
    the BH. The situation in Run~E was even more extreme, as there was no disc formation or BH
    accretion. To form a single disc requires the cloud to move on an orbit 
    that passes close enough to the BH that the cloud be tidally disrupted by it. 
    From these two 
    runs it is also apparent that for our cloud's geometry to allow the formation of multiple
    structures the orbit must be radial to the point that the cloud passes within a distance from
    the BH less than its own semi-major axis $h$.
    
    Alternatively, it may be stated that the angle from the BH between the cloud's midplane and one
    of its tips, when at closest approach to the BH, is the maximum possible misalignment. Thus a 
    head-on collision would produce a
    misalignment of $90\degree$ (although if the flows were symmetric, the angular momentum would
    be cancelled and the streamers would not form), while a distant pass
    would form a single disc only, and then only if tidal forces were still able to disrupt it.
    That such nearly radial orbits are required may be the largest caveat to the formation
    of multiple stellar discs by this method.
    
\begin{figure*} \centering
      \begin{tabular}{ccc}
        \includegraphics[height=4.5cm]{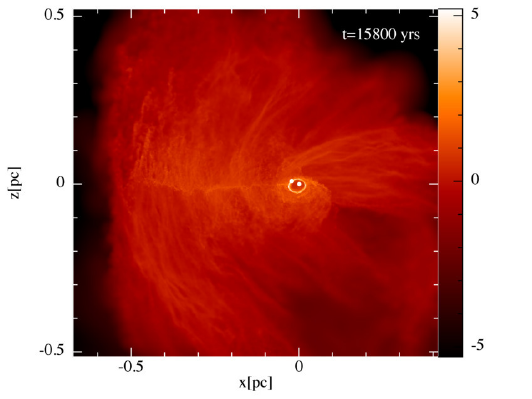} &
        \hspace{-0.5cm}\includegraphics[height=4.5cm]{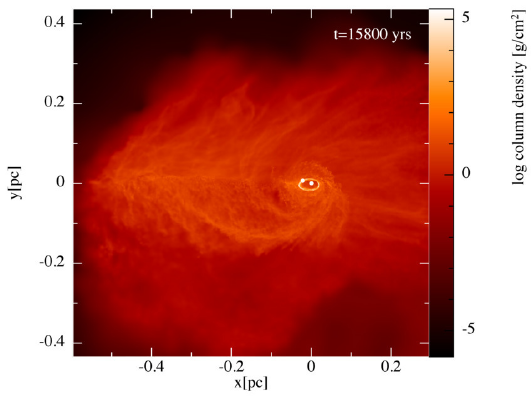} &
        \hspace{-0.3cm}\includegraphics[height=2.8cm]{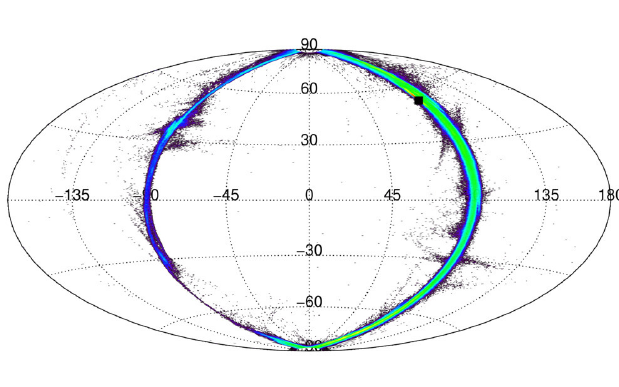} \\
        \includegraphics[height=4.5cm]{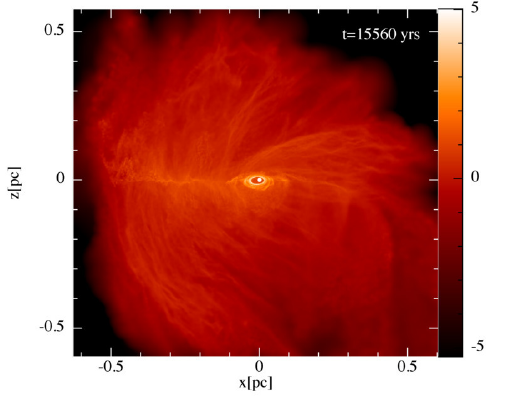} &
        \hspace{-0.5cm}\includegraphics[height=4.5cm]{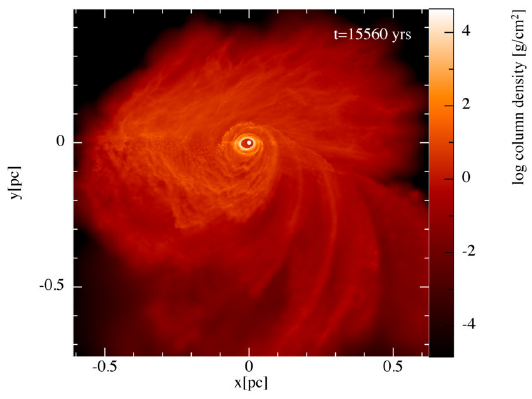} &
        \hspace{-0.3cm}\includegraphics[height=2.8cm]{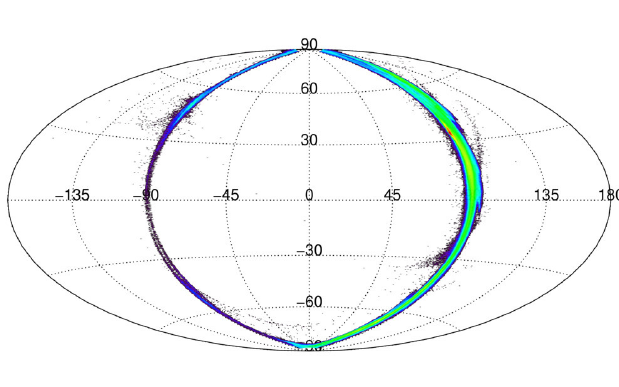}\\
        \includegraphics[height=4.5cm]{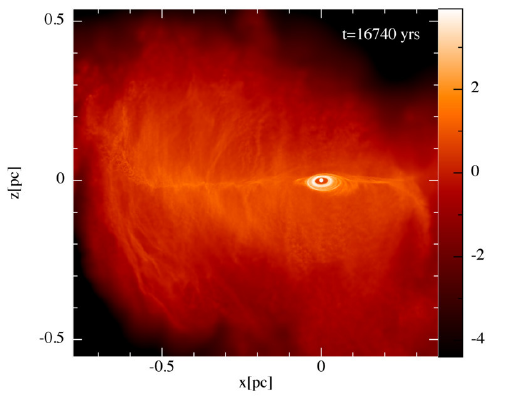} &
        \hspace{-0.5cm}\includegraphics[height=4.5cm]{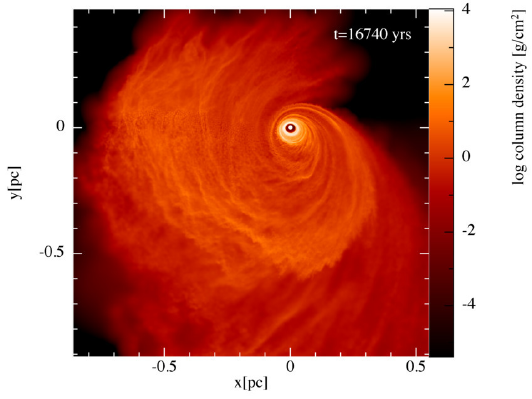} &
        \hspace{-0.3cm}\includegraphics[height=2.8cm]{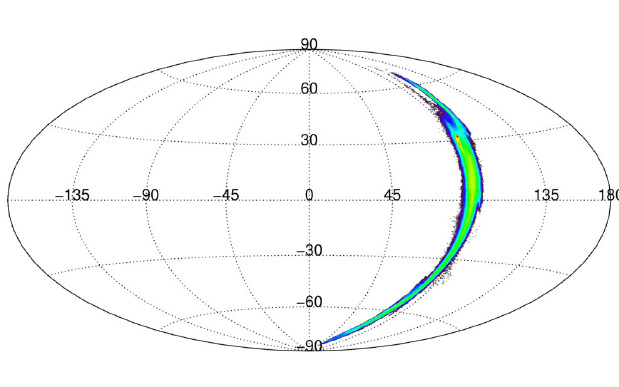}\\
        \includegraphics[height=4.5cm]{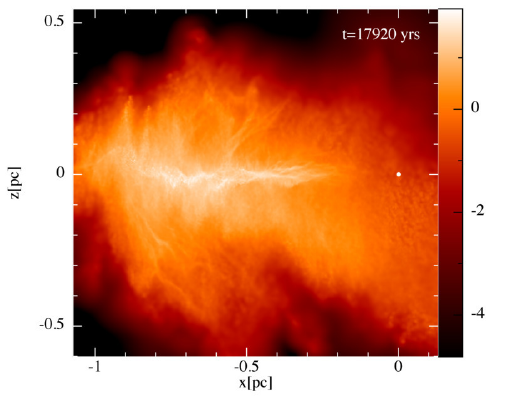} &
        \hspace{-0.5cm}\includegraphics[height=4.5cm]{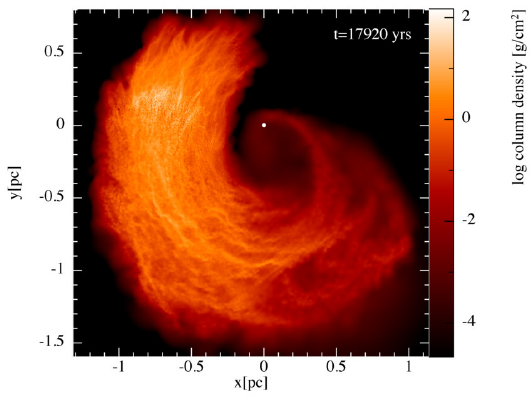} &
        \hspace{-0.3cm}\includegraphics[height=2.8cm]{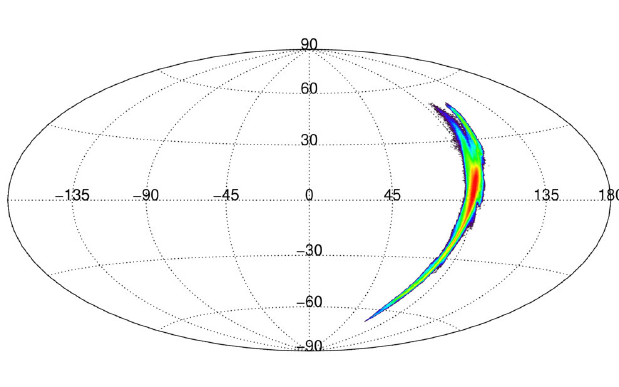}
      \end{tabular}
      \caption{$xz$ and $xy$ column densities, and angular momentum spread of all gas in Hammer 
        projection (see Figure~\ref{standard_menc_GC2D121_angmom}) of Runs~H5, H10, H20 and H40. The
        first three are at their end-points, and Run~H40 is shown at a time which allows comparison.
        From top to bottom, the cloud's tangential velocity $v_y$ was $5$, $10$, $20$, and 
        $40\,\rmn{km}\,\rmn{s}^{-1}$. The most direct cloud-BH interaction, with 
        $v_y=5\,\rmn{km}\,\rmn{s}^{-1}$, possessed the greatest spread in angular momentum. As $v_y$
        increased, the spread in angular momentum shrank, and the flow consolidated to form a single
        disc, as seen in the $v_y=20\,\rmn{km}\,\rmn{s}^{-1}$ case. When the tangential velocity was
        $40\,\rmn{km}\,\rmn{s}^{-1}$, only a very small portion of the cloud was captured.}
      \label{infall150_runs}
    \end{figure*}

\begin{figure*} \centering
      \begin{tabular}{ccc}
        \includegraphics[height=4.5cm]{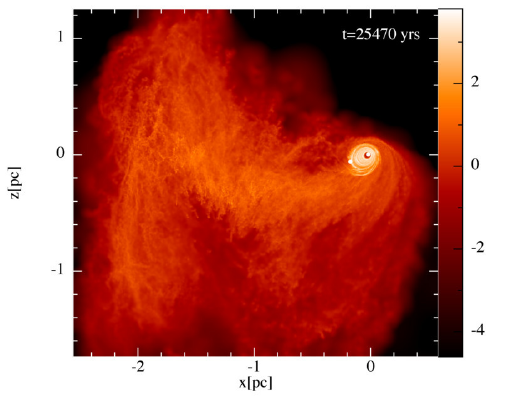} &
        \hspace{-0.5cm}\includegraphics[height=4.5cm]{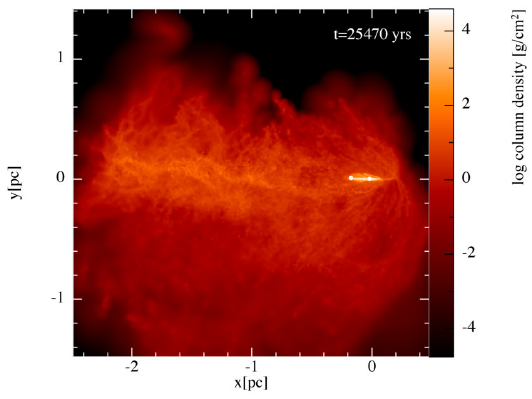} &
        \hspace{-0.3cm}\includegraphics[height=2.8cm]{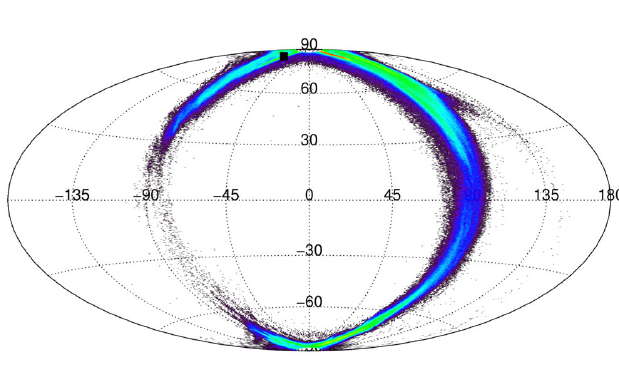} \\
        \includegraphics[height=4.5cm]{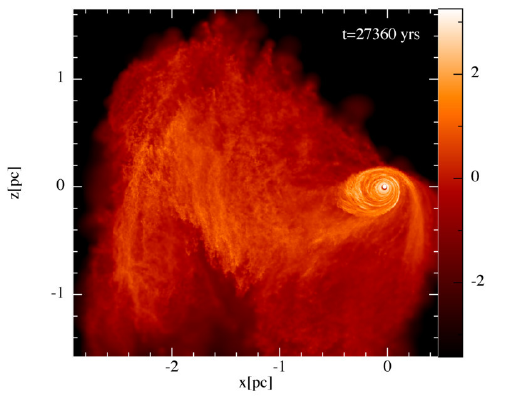} &
        \hspace{-0.5cm}\includegraphics[height=4.5cm]{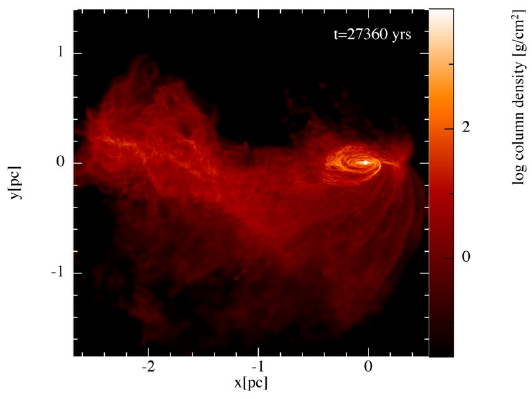} &
        \hspace{-0.3cm}\includegraphics[height=2.8cm]{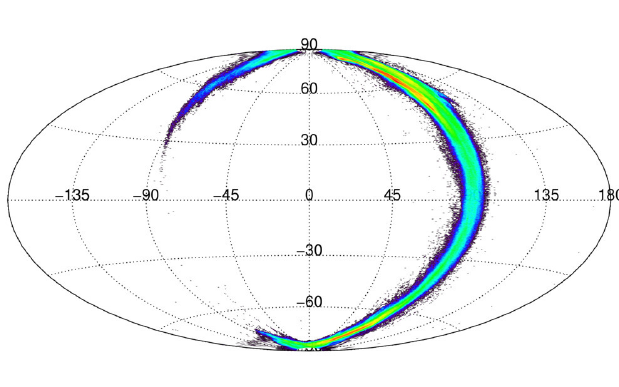}\\
        \includegraphics[height=4.5cm]{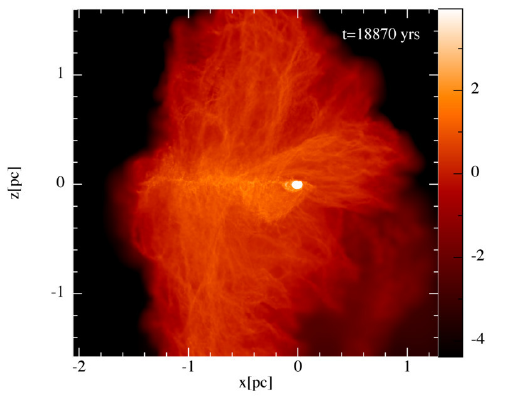} &
        \hspace{-0.5cm}\includegraphics[height=4.5cm]{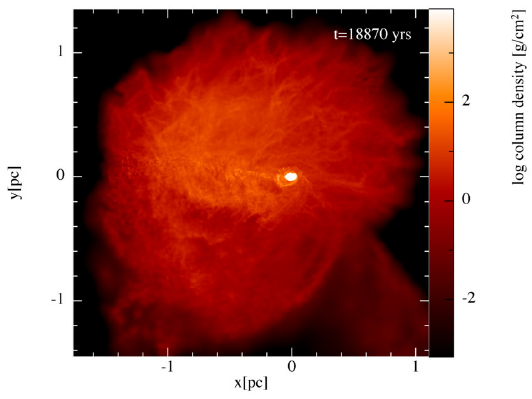} &
        \hspace{-0.3cm}\includegraphics[height=2.8cm]{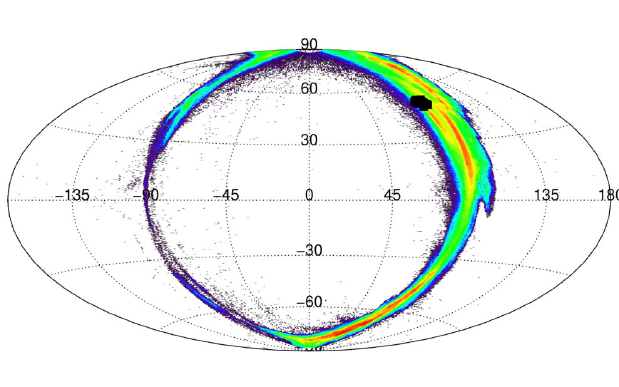}\\
        \includegraphics[height=4.5cm]{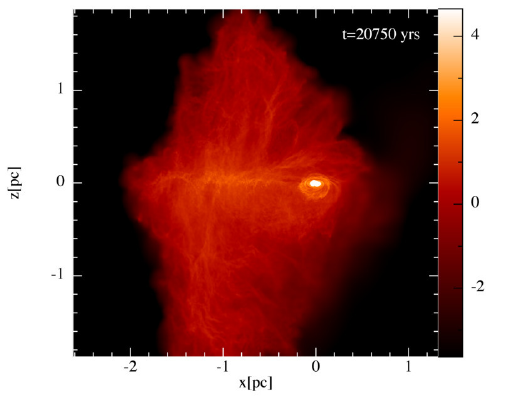} &
        \hspace{-0.5cm}\includegraphics[height=4.5cm]{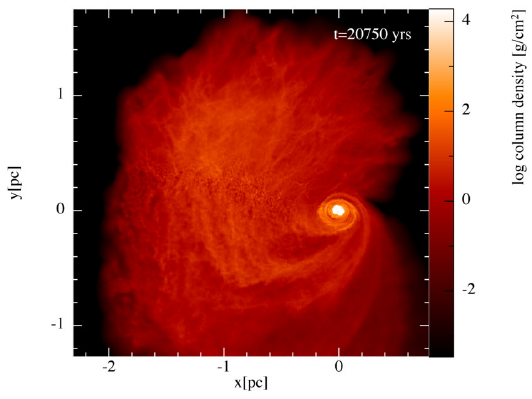} &
        \hspace{-0.3cm}\includegraphics[height=2.8cm]{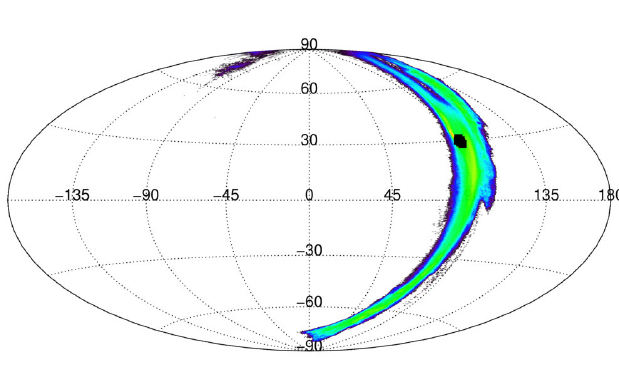}
      \end{tabular}
      \caption{The end states of Runs~I5 to I40 are shown here, with one simulation per row. As in
        Figure~\ref{infall150_runs}, the first plot on each row shows column densities in the 
        $xz$-plane, the second in the $xy$-plane, and the final shows the spread in the orientation
        of the particles' angular momenta on the sky. As the initial tangential velocity $v_y$
        increased, it can be seen that the primary disc which formed rotated farther into the 
        $xy$-plane, as was seen with Runs~H5 to H40. Runs~I20 and I40 were unable to progress as
        far, and so do not contrast so well with the first two runs. The increased cloud size has 
        broadened the band of angular momentum orientations seen in the third column. Again,
        increasing $v_y$ has led to the gas motions around the BH growing more aligned, though
        the enlarged cloud has noticeably helped to preserve the large angular spread of angular
        momentum when compared to Runs~H5 to H40.}
      \label{infall150_density_runs}
    \end{figure*}

    The H and I runs all used the much higher infall speed $v_x = -150\,\rmn{km}\,\rmn{s}^{-1}$
    predicted by the BH and cluster potential model. Each run's name includes a number, $5$, $10$,
    $20$, or $40$, indicating the tangential speed $v_y$ in $\rmn{km}\,\rmn{s}^{-1}$. While the 
    H-runs used 
    $M_\rmn{cl} = 10^4 M_\odot$, the clouds in the the I-runs had $10^5 M_\odot$ and axes
    twice as long as those of Run~A's cloud. This meant the I-runs' clouds engulfed the
    BH even with $v_y = 40\,\rmn{km}\,\rmn{s}^{-1}$. Column density and angular momentum density 
    plots are shown for these two sets of runs in Figures~\ref{infall150_runs} and 
    \ref{infall150_density_runs}. 
    
    With the same turbulent velocities, these clouds formed the same overdensity in the cloud centre
    as in Run~A from which it formed the disc. Positioned above the cloud's midplane, with low $v_y$
    the disc formed not in the orbital $xy$-plane
    but almost in the $xz$-plane. As the tangential velocity $v_y$ was increased, the disc rotated 
    towards the $xy$-plane, clearly visible in Figures~\ref{infall150_runs} and 
    \ref{infall150_density_runs}. This was reflected by the angular spread of angular momentum 
    consolidating towards the $z$-axis with increasing $v_y$.
    
    Infall was however so fast during these runs that asymmetries could not grow significantly, and
    shocking took place on the BH's far side. No misaligned streamers such as those described in
    the previous section could form, except in the case Run~I10. Gas from the top half of the cloud
    was able to pass downwards on its orbit and avoided shocking. As it reached the BH it was 
    sheared
    further to form a disc on the scale $\approx 0.2 \,\rmn{pc}$. This had not however depleted the
    gas reservoir, which still possessed a large range in angular momentum. Hence, as a different
    subset of the gas fell inwards and was itself sheared, it formed a second structure of similar
    density to the inner disc, but positioned slightly outside and oriented $17\degree$ out of
    plane.
    
\begin{figure} \hspace{-0.5 cm}
      \includegraphics[scale=0.46]{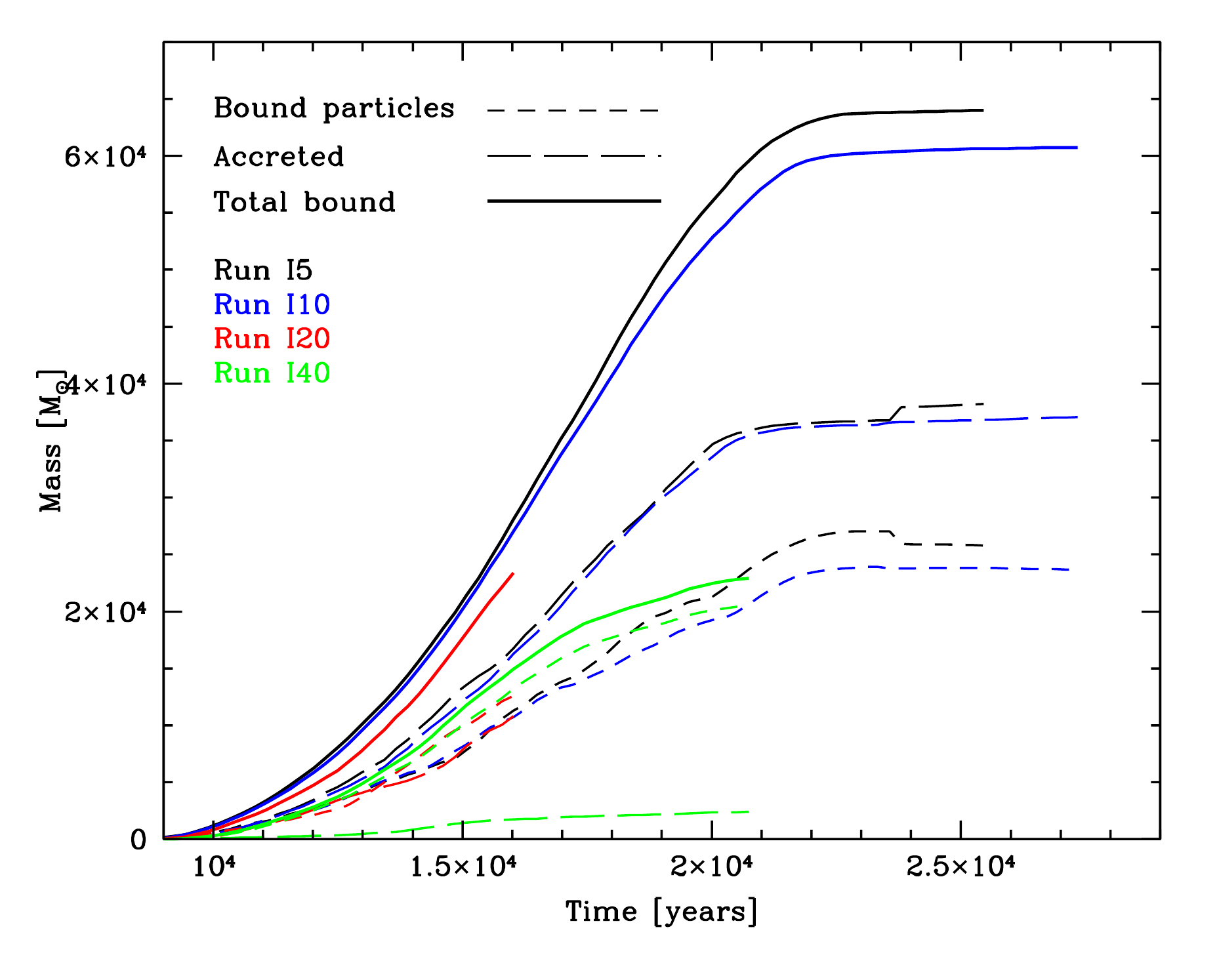}
      \caption{Mass of gas bound to within a distance of  $1\,\rmn{pc}$ from the BH 
        (short dashes) and the mass accreted to the BH (long dashes) have been plotted here against 
        time for the I-simulations. The thick solid lines show the sum of the two --
        this is the total mass bound to within $1\,\rmn{pc}$. The total mass dropped 
        from $6.4 \times 10^4 M_\odot$ to $2.2\times 10^4 M_\odot$ as the initial tangential 
        velocity $v_y$ increased from $5$ to $40\,\rmn{km}\,\rmn{s}^{-1}$. The majority of this is
        due to a decrease in mass accreted to the BH sink particle. The kinks seen in the I5 lines
        (black) at $24000\,\rmn{yrs}$ came about as at that time very short timestep gas particles
        close to the BH were forcibly accreted to it in order for it to run slightly longer.
        By the end of Runs~I5 and I10, roughly twice as much material was accreted to the BH
        sink as remained bound in the gas phase. In both cases the high infall speed (reducing
        structure formation time) and low $v_y$ (reducing $z$-component angular momentum)
        brought about extremely effective shocking behind the BH, allowing rapid accretion of low
        angular momentum material.
        }
      \label{Iruns_tot_captured_mass}
    \end{figure}

    The mass in gas that orbits close to the BH is an upper limit to the mass of any stellar disc.
    Figure~\ref{Iruns_tot_captured_mass} shows the time evolution of the mass in gas particles bound
    to within $1\,\rmn{pc}$ and the mass accreted to the BH for all of the I-runs. (We
    remind the reader that the BH sink particle represents not only the physical BH but also its 
    unresolved inner accretion disc.) While the simulations end at different times, there is a 
    clear trend of decreasing total mass (solid lines) with increasing initial cloud tangential 
    velocity $v_y$. It
    is in accreted mass (long dashes) that the majority of the decrease occurs; the final mass 
    accreted to the BH sink varied from almost $4\times 10^4 M_\odot$ for I5 and I10, to only a few 
    thousand $M_\odot$ for I40. When $v_y$ was low, the $z$-component of the angular momentum
    was reduced. Thus once shocking had taken place on the far side, itself enhanced by the high
    infall speed, very little angular momentum remained. Changes in the mass of bound gas (short 
    dashes) are much smaller and indicate final values of $2$ to $3 \times 10^4 M_\odot$ 
    for these four runs. As such, with low $v_y$ (Runs~I5 and I10) the shocking was so effective
    at cancelling out angular momentum that more mass was accreted to the BH than remained bound
    in the gas phase. For Runs~I20 and I40, this had reversed with more being bound and less
    accreted.
    
    Only particles within $0.02\,\rmn{pc}$ of the sink particle representing the BH could be
    accreted. Furthermore, they were required to have a low enough angular momentum to form a
    circular orbit at that radius. Assuming a particle's angular momentum remains constant during
    infall, its circularisation radius is $R_\rmn{circ} \propto l^2$ where $l$ is the specific
    angular momentum. The low angular momentum needed to circularize within the BH's accretion
    radius was available to fewer particles when the cloud's initial $v_y$ was higher. Meanwhile, 
    those with larger angular momentum were still able to circularize within $1\,\rmn{pc}$.
  
\section{Star formation} \label{s:sf}
  Although the focus of this paper is placed on the gas dynamics of a cloud infalling towards a
  massive BH, we here take time to discuss the star formation we observed in our simulations.
    
  \subsection{Star formation physics}
    We included an approximation of radiative cooling via the method of \citet{FO09}. Before 
    fragmentation could even begin to occur, the gas disc had to become cool enough that thermal 
    pressure would not impede collapse. This is normally described by the Toomre 
    Q-criterion which also ensures that rotational energy cannot form a barrier against 
    self-gravity.
    
    A fragment's ability to collapse and form a star is dependent on the
    ability of the gas to cool quickly (\citealt{GA01}; \citealt{REA03}). As the parcel of gas
    contracts, it heats up, which would halt the collapse due to the increase of thermal
    pressure. If the gas can cool quickly, this should not be a problem and the
    collapse can continue. On the other hand if the cooling is slow the shear in the disc will
    disperse the fragment before it can collapse. This means that the cooling timescale must be
    shorter than a few times the disc's rotational period \citep{GA01}. In discs
    around SMBHs the cooling time is typically very short inside $R=0.1\,\rmn{pc}$ (\citealt{LB03};
    \citealt{LE07}; \citealt{NCS07}; \citealt{AL08}).
    
  \subsection{Sinks in the simulations}
\begin{figure}
      \includegraphics[scale=0.46]{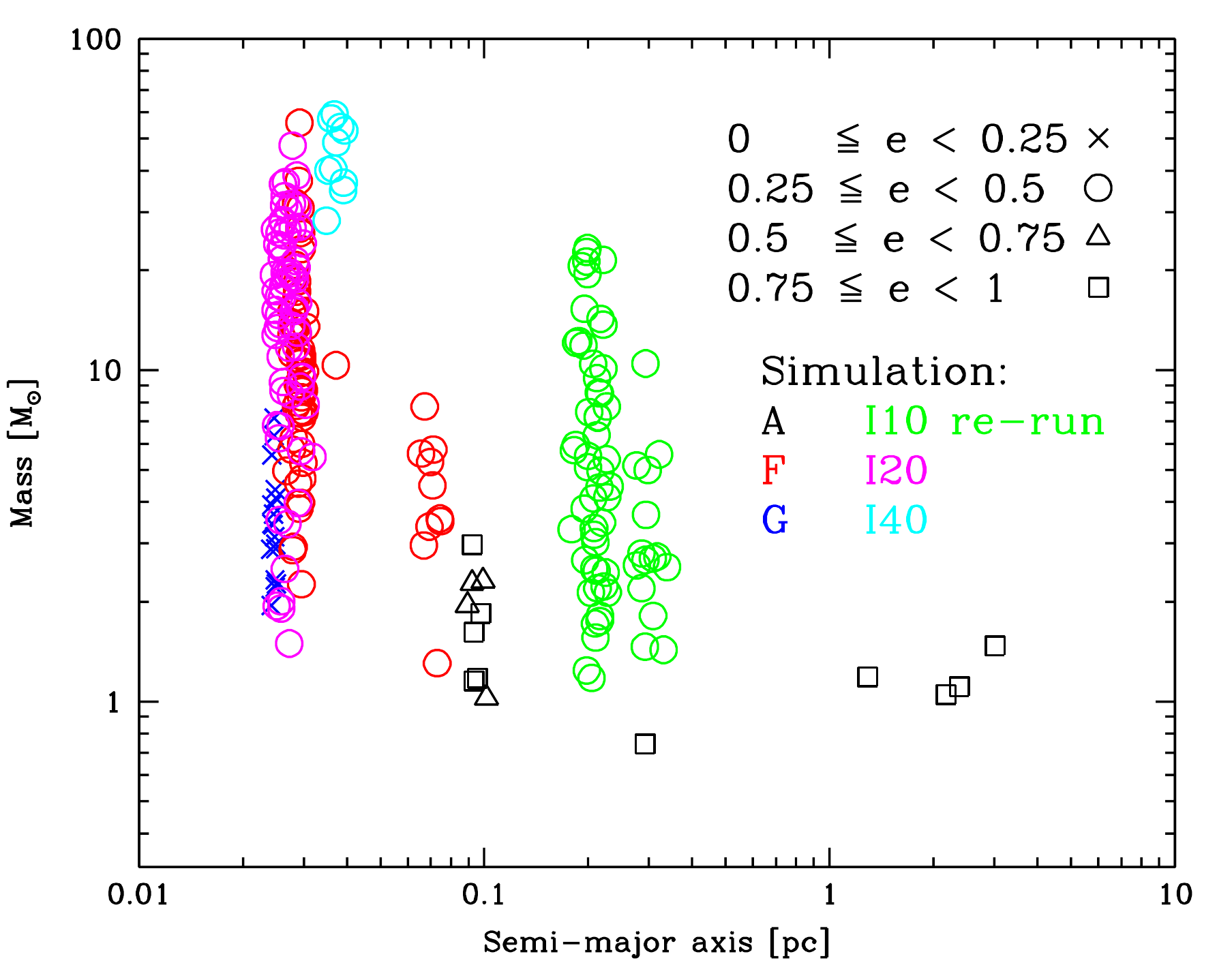}
      \caption{The masses of sinks from several simulations are plotted against their semi-major
        axes. The symbols' colours show which run they were from, and the shape indicates which
        eccentricity range that sink inhabited. Most simulations' sinks are seen to inhabit 
        well-defined regions. The lowest eccentricity sinks were only found at the lower end of the
        semi-major axis range as a direct result of the cancellation of angular momentum by gas 
        during shocking and torquing. The highest mass sinks were also found at low semi-major axes.
        Higher eccentricity sinks orbited at larger distances, and had lower
        masses. The inner and outer populations of sinks in Run~F and the re-run of I10 (where the
        outer was $17\degree$ out-of-plane from the inner) can be clearly distinguished. Those in
        Run~A at the largest distances were formed during infall and did not belong to the disc.
        Perhaps most noticeably, eccentricities in the range of $0.25$ to $0.5$ were by far the
        most common, and covered almost the complete range of masses and semi-major axes.}
      \label{mass_vs_sma_eccen}
    \end{figure}

    It was not uncommon for sinks to form in the cloud during slow infall 
    ($v_y = -41.5\,\rmn{km}\,\rmn{s}^{-1}$). These were either unbound or on orbits with 
    eccentricities approaching unity. Our interest however lay with star formation within a disc or
    streamer, as demonstrated in earlier numerical experiments (\citealt{NCS07}; \citealt{BR08}; 
    \citealt{HN09}; \citealt{AL11}; \citealt{MA12}). The masses, semi-major axes and eccentricities
    of the sinks we discuss in this section are shown in Figure~\ref{mass_vs_sma_eccen}.
    
    By the end of Run~A at $t = 28300\,\rmn{yrs}$, nine sinks had formed in the disc and
    followed orbits with semi-major axes $a \approx 0.095\rmn{pc}$ and eccentricities 
    $e \approx 0.74$. Forming shortly before the simulation's end, they did not have enough time to
    accrete to high masses and ranged from one to three $M_\odot$. While the number of sinks formed
    in this simulation is well below the observed number of more than a hundred, it is likely that
    that number would have increased. In the integration step during which the simulation ended,
    eighteen further sinks were created, more than doubling the total; the step was only half
    completed. Indeed the rate of sink formation was accelerating. Thus our sinks are individually
    realistic, but our sample incomplete.
    
    \begin{figure} \centering
      \includegraphics[angle=-90,scale=1.3]{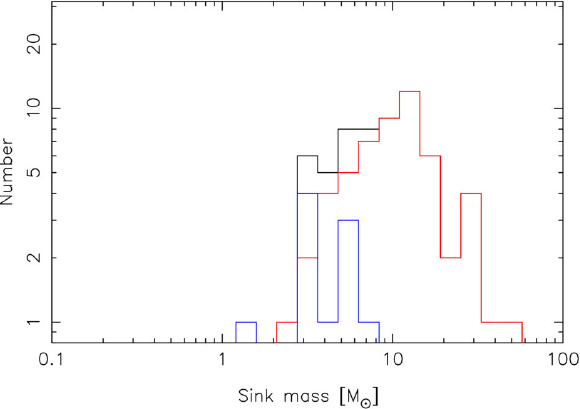}
      \caption{Mass functions for the two populations of sinks seen at $25000\,\rmn{yrs}$ in Run~F.
        The red line shows the mass function for all sinks with semi-major axes 
        $a \leq 0.05\,\rmn{pc}$, and the blue line those with $a > 0.05 \,\rmn{pc}$. The black
        line shows the total. While both populations contain sinks which would be considered
        high mass, those in the outer population have had less time to accrete and so do not reach
        the extremely high masses of those in the inner population. No masses are below 
        $\sim 0.9 M_\odot$, equivalent to $30$ gas particles.}
      \label{mf_density_menc_GC2D107_2pops}
    \end{figure}
    
    The runs using the cloud with mass $M_\rmn{cl} = 10^5 M_\odot$ (F and the I runs) were able
    to form a greater number of sinks. By Run~F's end at $t=25000\,\rmn{yrs}$, sixty-four sinks were
    orbiting the BH. These could be split into two groups separated by semi-major axis. The first, 
    consisting of fifty-four sinks, had semi-major axes of $a \approx 0.03\,\rmn{pc}$ and 
    eccentricities $e \approx 0.33$. These began to form at $t = 22070\,\rmn{yrs}$. The second 
    population, consisting of ten sinks, began to form at $24520\,\rmn{yrs}$ and had 
    $a \approx 0.07\,\rmn{pc}$ and $e \approx 0.43$. The mass function of the sinks at the 
    simulation's end are shown in Figure~\ref{mf_density_menc_GC2D107_2pops}. The sinks were
    able to accrete quickly; by this point at the end the most massive sink had $55.8 M_\odot$, and
    the mean mass was $11.6 M_\odot$.
    
    The sinks' semi-major axes and eccentricities are consistent with those observed, though the 
    former lie at the lower end of the reported range. In this run, the higher gas mass produced 
    greater torques and shocks which were more effective in reducing the angular momentum of the 
    captured material. Thus the gas disc and stellar orbits were smaller and more circular than
    those seen in Run~A.
    
\begin{figure} \centering
      \includegraphics[scale=1.6]{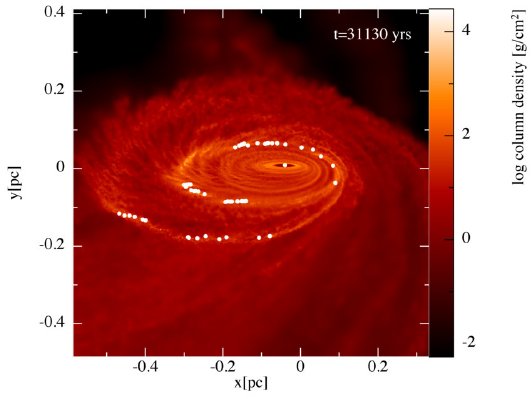} \\
      \includegraphics[scale=1.6]{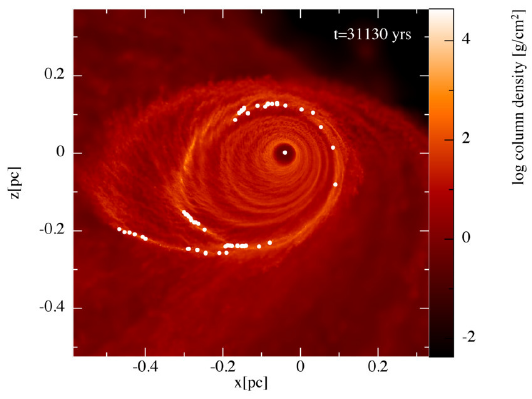} \\
      \includegraphics[scale=1.6]{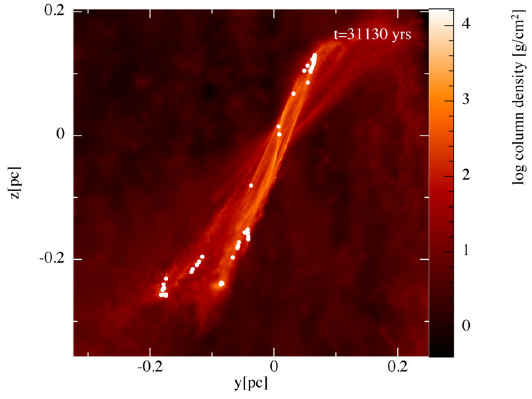}
      \caption{Column densities for the re-run of I10 in the $xy$, $xz$ and $yz$ planes at the simulation's 
        end at $t = 31130\,\rmn{yrs}$. By this point the disc and gas streamers feeding it had grown
        dense enough that sink formation was underway. The streamers were oriented $17\degree$
        out-of-plane from the disc, resulting in two stellar systems separated by this angle.}
      \label{ColDens_infall15_tang10_density_new}
    \end{figure}

    No sinks formed in misaligned streamers in simulations with slow infall
    $v_x = -41.5\,\rmn{km}\,\rmn{s}^{-1}$. We performed a re-run of Run~I10, where a streamer
    was formed at $17\degree$ to the BH accretion disc. Here we allowed the BH to accrete
    all material within $0.02\,\rmn{pc}$ without test and sinks to form with ratio of
    thermal to gravitational energy of one, as opposed to $0.5$ which is otherwise used. With these
    changes the simulation could run to $t = 31130\,\rmn{yrs}$, and the misaligned streamer was able
    to become dense enough that a second population of sinks formed $17\degree$ from the first. The
    simulation at its end is shown in Figure~\ref{ColDens_infall15_tang10_density_new}.
 
\begin{figure} \centering
      \includegraphics[angle=-90,scale=1.3]{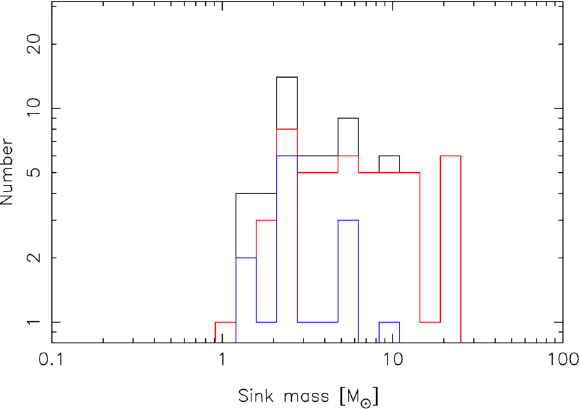}
      \caption{Mass function for the re-run of I10 at $t = 31130\,\rmn{yrs}$. The red and blue lines
        represent the individual mass functions for the disc ($a \approx 0.2\,\rmn{pc}$) sinks and
        the streamer ($a \approx 0.3\,\rmn{pc}$) sinks, respectively. The underlying black line
        shows the MF for the total population. The outer population, which formed later on, consists
        of lower mass sinks, while the inner population formed earlier and so had time to accrete
        to the high masses we see here.}
      \label{mf_infall150_tang10_density_new_GC2D133_2pops}
    \end{figure}

    The fifty-two sinks in the first population had semi-major axes of $\approx 0.2\,\rmn{pc}$ and 
    eccentricities of $\approx 0.4$; the fifteen in the second population, formed in the streamer, 
    had semi-major axes of $\approx 0.3\,\rmn{pc}$ and eccentricities of $\approx 0.45$. Mass 
    functions are shown in Figure~\ref{mf_infall150_tang10_density_new_GC2D133_2pops}. While
    the outer misaligned sinks are at lower masses, both populations contain sinks which have been
    allowed to accrete to high masses.
    
    The MFs shown (Figures~\ref{mf_density_menc_GC2D107_2pops} for Run~F and 
    \ref{mf_infall150_tang10_density_new_GC2D133_2pops} for the re-run of I10) do not resemble a
    standard 
    \citet{SA55} IMF, being flat in logspace from one to ten $M_\odot$. The number of sinks in each
    bin for Run~F's inner population actually increases, peaking in the range of tens of solar
    masses before falling again. The observed IMF is likewise weighted towards the higher end
    \citep{BA10}, apparently a violation of a `global' Salpeter-like IMF. Simulations suggest this
    is due to high gas
    temperatures and preferential formation of high-mass fragments due to tidal forces \citep{BR08}.
    Variable tidal forces experienced throughout an eccentric disc such as
    those in our simulations were also found by \citet{AL08} to increase the fragment mass when
    compared to those in circular discs. It is important to emphasise that our plots are not a
    numerically-derived initial mass function, as the sinks are still accreting. 
    This does however show that that our sinks have been able to grow to tens of solar masses
    within only a few thousand years.
    
    \begin{figure} \centering
      \includegraphics[scale=0.4]{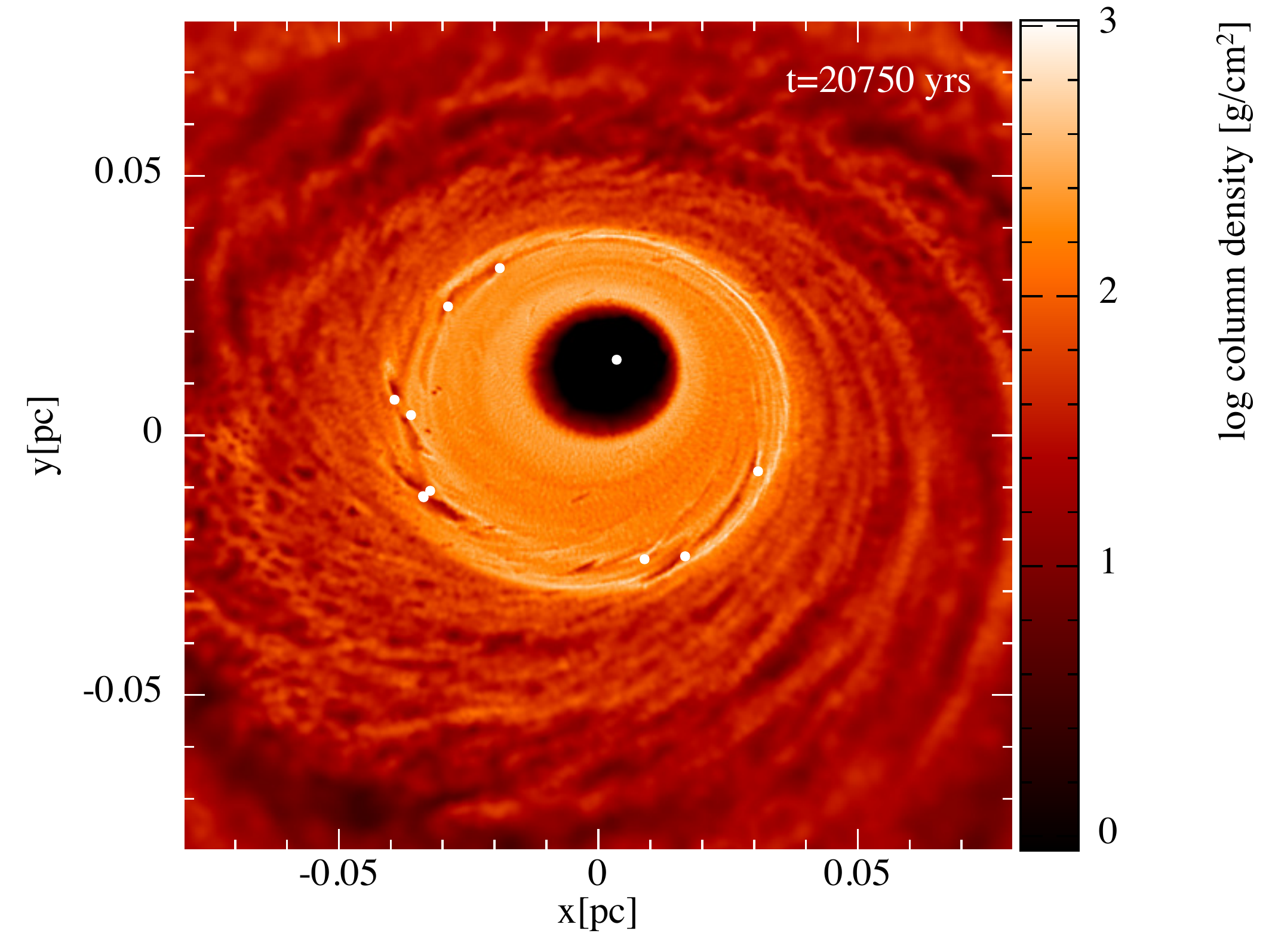}
      \caption{Column density in the $xy$-plane for Run~I40 at its end-time of $20750\,\rmn{yrs}$.
        White dots show the locations of sink particles. The high initial tangential velcocity
        of $40\,\rmn{km}\,\rmn{s}^{-1}$ has suppressed the formation of any misaligned gas flows. 
        Shocked gas has formed a small, low-eccentricity disc. The ten sinks within it have
        $a\approx 0.037\,\rmn{pc}$ and $e\approx 0.3$. Gaps in the disc mark the regions where the
        sinks have been accreting from the gas, raising them to several tens of solar masses within
        $5000\,\rmn{yrs}$.}
      \label{ColDens_infall150_tang40_density_GC2D089_xy_zoom}
    \end{figure}
    
    The gaseous discs which formed within Run~I20 and I40 were also able to form a substantial
    number of sinks. Considerable shocking had taken place within the gas, and so the disc and
    stellar orbits were small with semi-major axes $a\approx 0.027\,\rmn{pc}$ and eccentricities 
    $e\approx 0.4$. The mass function resembled that of the inner population of Run~F (red line, 
    Figure~\ref{mf_density_menc_GC2D107_2pops}), but it was actually shifted even further towards
    high masses, peaking at $\sim 30 M_\odot$. Run~I40 formed only ten sink particles (see 
    Figure~\ref{ColDens_infall150_tang40_density_GC2D089_xy_zoom}), with 
    $a\approx 0.037\,\rmn{pc}$ and $e\approx 0.3$. Despite their small number, their masses were 
    very large -- the lowest sink mass was $28 M_\odot$, and the highest $59 M_\odot$. Noticeable
    openings in the disc show the regions from which they had been accreting.
    
  \subsection{Subsequent star formation}
    By the time our simulations ended, no sink particles had been created in streamers at an angle
    beyond $17\degree$ from the principle gas disc. Our simulations were only able to run for a
    few tens of thousands of years. Given the physical conditions at the end of the simulations we
    expect further fragmentation to occur.
       
\begin{figure} \centering
      \includegraphics[scale=1.3]{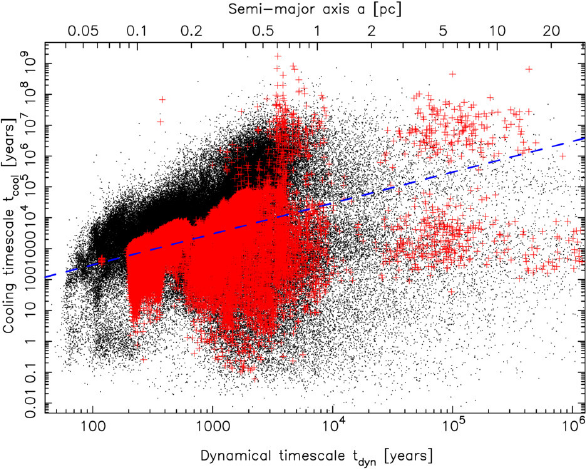}
      \caption{Cooling and dynamical timescales (with corresponding semi-major axis) for all
        particles in Run~A at $t=28300\,\rmn{yrs}$. Those marked with red crosses are above the
        tidal density; the dashed line marks $t_\rmn{cool} = 3t_\rmn{dyn}$. Gas below this 
        line can fragment according to \citet{GA01}.}
      \label{tcooldyn_standard_menc_GC2D121}
    \end{figure}

    \citet{GA01} found that 
    only gas which is able to cool over a timescale $t_\rmn{cool}$ shorter than three times the 
    dynamical timescale $t_\rmn{dyn}$ may fragment -- see also \citet{REA03} and \citet*{RLA05}). 
    Figure~\ref{tcooldyn_standard_menc_GC2D121} shows the cooling and dynamical timescales for
    Run~A at its end ($t=28300\,\rmn{yrs}$). A wide dispersion of $t_\rmn{cool}$ over several
    orders of magnitude above and below the critical $t_\rmn{cool} = 3 t_\rmn{dyn}$ line can
    be seen throughout. Within $a \approx 0.1\,\rmn{pc}$, the majority of the gas had $t_\rmn{cool}$
    between $\sim 100$ and $\sim 10^4\,\rmn{yrs}$. This gas should be able to cool sufficiently to 
    form additional sinks.
    
    That gas for which the density exceeded the tidal value is also marked in 
    Figure~\ref{tcooldyn_standard_menc_GC2D121}. Of the $17904 M_\odot$ 
    of gas which had not been accreted to the BH or formed sink particles, $7353 M_\odot$ had 
    $t_\rmn{cool} \le 3 t_\rmn{dyn}$; of that, $1689 M_\odot$ exceeded the tidal density. 
    This self-gravitating gas with a short cooling time may
    fragment and collapse. If we then assume that all this gas was destined to form stars, it 
    suggests a final star formation efficiency of $8.5$ per cent for the initial cloud of 
    $2\times 10^4 M_\odot$.
    
\begin{figure} \centering
      \includegraphics[scale=1.3]{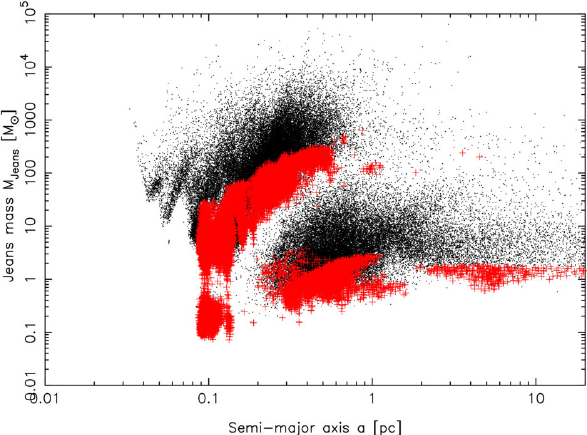} \\
      \includegraphics[scale=1.3]{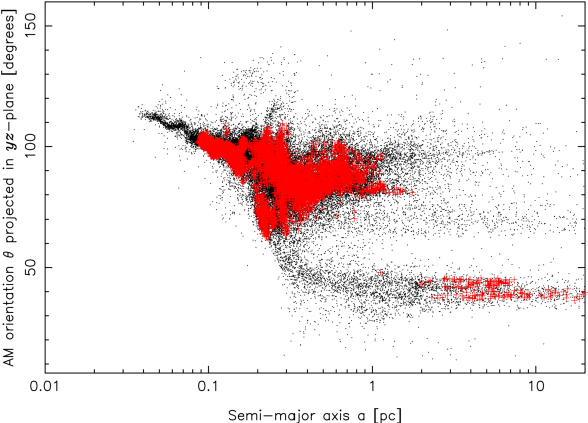}
      \caption{Here we compare both the Jeans mass $M_\rmn{Jeans}$ and orbital inclination 
        $\theta$ with the semi-major axis $a$ of gas in Run~A at $t=28300\,\rmn{yrs}$. Here we 
        define the orbital inclination to be $\theta = \rmn{tan}(l_z/l_y)$ i.e. the angle between 
        the simulation $y$-axis and the projection of the angular momentum vector into the 
        $yz$-plane. Since the $x$-component of the angular momentum is small 
        (Figure~\ref{standard_menc_GC2D121_cumulative_mass_angmom}), $\theta = 0\degree$ means the 
        gas is orbiting almost in the $xz$-plane, and $90\degree$ in the $xy$-plane. We only plot
        that gas which fulfilled the Gammie criterion for fragmentation in 
        Figure~\ref{tcooldyn_standard_menc_GC2D121}, that is to say lay beneath the line. Red 
        crosses again mark gas denser than the local tidal value. 
        The gas in the disc follows orbits with $a \lesssim 1\,\rmn{pc}$. Jeans masses range from
        $0.1$ to $300 M_\odot$. The streamer can be seen in the second plot at 
        $\theta \approx 40 \degree$ and extending to large semi-major axes. The first plot shows 
        that Jeans masses within it are small, reaching at most $2 M_\odot$.}
      \label{fig:sink_predictions}
    \end{figure}

    We can also estimate the properties of the stars which will form. We take all the gas
    from Figure~\ref{tcooldyn_standard_menc_GC2D121} that has a short enough $t_\rmn{cool}$ for 
    collapse and plot the Jeans
    mass against semi-major axis; this can be seen as the first plot in Figure~\ref{fig:sink_predictions}.
    Again we highlight those particles exceeding the tidal density. We also calculate the angle
    between the $y$-axis and the projected $yz$-angular momentum for this gas; this is shown in
    the second plot of Figure~\ref{fig:sink_predictions}. The gas on smaller orbits has Jeans masses
    varying from $0.1$ to $300 M_\odot$. The second plot shows that this gas also covers about
    $50 \degree$ in orbital inclination, centred roughly around $\theta \approx 90\degree$ i.e.
    angular momentum vector aligned with the $z$-axis. This reflects that by the simulation's end
    some gas was still in the process of settling into the disc plane.
    
    The gas on larger orbits can be seen in second plot of Figure~\ref{fig:sink_predictions} to
    be restricted to an orbital orientation of $\theta \approx 40 \degree$ -- this is the
    misaligned streamer. While it seems that stars should form here over time, the Jeans masses
    indicated previously for this gas are much lower, ranging from $0.5$ to $2 M_\odot$. Also, the
    semi-major axis for streamer gas extends to distances of $10\,\rmn{pc}$. The mass of gas in the 
    streamer which may form stars is very low however, at only $21 M_\odot$.
    
\section{Conclusions} \label{s:conclusions}
  Recent observations
  indicate the presence of two misaligned discs of massive stars around Sagittarius A* 
  (\citealt{PA06}; \citealt{BA09}). These stars all formed at roughly the same time, somewhere
  between 6 and 10 million years ago. If these stars were formed via \emph{in situ} disc 
  fragmentation \citep{LB03}, two misaligned discs or streamers are required. In this paper we
  have examined whether the infall of a single cloud to the BH to provide the star formation 
  material is a plausible scenario.
  
  the cloud need also have an impact parameter  which is smaller than the largest extent of the 
  cloud. In addition, significant sub structure need exist within the cloud, such as produced here 
  by turbulence, to avoid symmetric shocks behind the black hole. This ensures sufficient angular
  momentum remains in the bound gas to form massive gaseous discs around the hole.
  
  We ran smoothed particle hydrodynamics (SPH) simulations of the infall of
  several clouds towards a $4\times 10^6 M_\odot$ black hole (BH). Crucially, the cloud was given an
  ellipsoidal geometry, with its major axis oriented perpendicular to its orbital plane. The cloud's
  orbit was highly radial to the point that the cloud marginally engulfed the BH. The combination
  of the two spread the distribution of gas particles' angular momenta about the BH
  over a large region of the sky. When the cloud reached the BH, its central regions were tidally
  sheared and captured to form a disc. The extremities flowed around the BH at large angles to one
  another and to the disc. If the cloud were to have uniform density, these flows would meet at the
  midplane and shock to remove their out-of-plane motion -- this is essentially the mechanism of 
  \citet{WYZ08} which allows the construction of small accretion discs. 
  The cloud's impact parameter was also required to be smaller than the largest extent of
  the cloud to ensure a sufficient spread in angular momentum. In addition, significant
  sub-structure was needed within the cloud, such as that produced here by turbulence, to avoid
  symmetric shocks behind the BH. This ensures sufficient angular momentum remains in the bound gas
  to form massive gaseous discs around the hole. These requirements indicate that the parameter
  space allowing the formation of a misaligned disc or streamer is somewhat limited.
  
  We also tested high infall speeds using $-150\,\rmn{km}\,\rmn{s}^{-1}$ while varying the 
  tangential speed from $5\,\rmn{km}\,\rmn{s}^{-1}$ to $40\,\rmn{km}\,\rmn{s}^{-1}$. The faster
  infall led to less structure forming by the time the cloud reached the BH. As angular momentum was
  modified with the tangential speed, we observed the gas flows rotating with increasing $v_y$. 
  In Run~I10, which used a $10^5 M_\odot$ cloud with $v_y = 10\,\rmn{km}\,\rmn{s}^{-1}$, a 
  streamer was able to form, though it was only misaligned from the primary disc by $17\degree$. In
  a re-run it was however able to progress far enough that sinks formed within both the central disc
  and the streamer.
  
  Across the simulations sink eccentricities varied from $0.15$ to $0.7$; \citet{BA09} reported the
  clockwise stars (the primary structure) to have a mean eccentricity of $0.35$. Semi-major axes of
  both the gas discs and any sink particles which formed within them were small. While some runs' 
  discs were as large as $0.3\,\rmn{pc}$, it was not uncommon for them to be smaller than 
  $0.1\,\rmn{pc}$. Meanwhile studies on the stellar discs has found them to inhabit a region between
  $0.05$ and $0.5\,\rmn{pc}$, or $1 - 10''$ (eg. \citealt{GE03}; \citealt{PA06}; \citealt{LU09}; 
  \citealt{BA09}).
  
  In Run~A, Jeans masses for gas above the tidal density in the inner disc ranged up to
  $\sim 100 M_\odot$. Streamer material orbited
  with semi-major axes of up to $10\,\rmn{pc}$. Only $21 M_\odot$ of gas within the streamer
  exceeded both the tidal density and was capable of fragmentation \citep{GA01}. Jeans masses were 
  much lower, between $0.5$ and $2 M_\odot$. As well as being separated from the main disc by
  $60\degree$, the gas capable of forming stars was only found on orbits with semi-major axes
  greater than $1\,\rmn{pc}$. In this instance, it seems unlikely that a misaligned disc such as 
  that observed will come about, but it could allow for the simulataneous formation of disc stars
  and additions to any present nuclear cluster. Given the sensitivity to initial conditions that
  we found, it is not impossible that another cloud would have formed stars on orbits similar to
  those observed. The other runs were capable of fragmentation with similar Jeans masses, but did
  not display the presence of the streamer.
  
  Depending on the cloud structure and orbit a variety of end configurations seem achievable and may
  provide a route to, for example, the counter-rotating gas discs of \citet*{NKP12}. We know from 
  previous work that the formation of large numbers of stars via fragmentation is viable 
  (\citealt{NCS07}; \citealt{BR08}; \citealt{HN09}; \citealt{AL11}; \citealt{MA12}). Together these
  indicate that a single gas cloud should be able to provide all the material for forming multiple
  stellar discs. 
  However, the requirements placed on the cloud's shape, structure and orbit as stated above
  may limit the effectiveness of the method.
 
\section*{Acknowledgements}
  WEL's PhD is supported by an STFC studentship from the Government of the United Kingdom. IAB
  acknowledges funding from the European Research Council for the FP7 ERC advanced grant project
  ECOGAL. WKMR acknowledges support from STFC grant ST/J001422/1.
  This work was supported by the Swedish Research Council (grants 2008--4089 and 2011--3991).
  The simulations presented here were run on UKMHD's Wardlaw cluster at St Andrews.
  Rainer Sch\"{o}del provided excellent advice regarding the nuclear stellar cluster which we used 
  to model its potential and for which we greatly thank him. Our column density plots were produced
  using Daniel Price's \textsc{splash} software \citep{PR07}. Finally, thanks go to the anonymous
  referee who provided many useful comments, allowing the paper to be improved.

\end{document}